\documentclass[12pt,preprint]{aastex}

\newcommand{\bvec}[1]{\textbf{#1}}

\shorttitle{Weak-lensing analysis of XMM2235}
\shortauthors{Jee et al.}

\begin{document}

\title{{\it HUBBLE SPACE TELESCOPE} WEAK-LENSING STUDY OF THE GALAXY CLUSTER XMMU J2235.3-2557 AT $Z\sim1.4$: \\A SURPRISINGLY MASSIVE GALAXY CLUSTER
WHEN THE UNIVERSE IS ONE-THIRD OF ITS CURRENT AGE\footnote{Based on observations made with the NASA/ESA {\it Hubble Space Telescope},
obtained at the Space Telescope Science Institute, which is operated by the Association of Universities for Research in Astronomy, Inc., 
under NASA contract NAS 5-26555, under program 10496 and 10698}}

\author{M. J. JEE\altaffilmark{2}, P. ROSATI\altaffilmark{3}, H. C. FORD\altaffilmark{4}, K. S. DAWSON\altaffilmark{5}, C. LIDMAN\altaffilmark{6},
S. PERLMUTTER\altaffilmark{7}, R. DEMARCO\altaffilmark{8}, V. STRAZZULLO\altaffilmark{9}, C. MULLIS\altaffilmark{10}, 
H. B\"{O}HRINGER\altaffilmark{11}, AND
R. FASSBENDER\altaffilmark{11}}

\begin{abstract}
We present a weak-lensing analysis of the $z\simeq1.4$ galaxy cluster XMMU J2235.3-2557,
based on deep Advanced Camera for Surveys images.
Despite the observational challenge set by the high redshift of the lens, we
detect a substantial lensing signal at the $\gtrsim8\sigma$ level. 
This clear detection is enabled in part by the high mass of the cluster, which
is verified by our both parametric and non-parametric estimation of the cluster mass.
Assuming that the cluster follows a Navarro-Frenk-White mass profile,
we estimate that the projected mass of the cluster within $r=1$ Mpc is
$(8.5\pm1.7)\times10^{14}$ $M_{\sun}$, where the error bar includes the statistical uncertainty of the shear profile,
the effect of possible interloping background structures, the scatter in
concentration parameter, and the error in our estimation of the mean redshift of the background galaxies.
The high X-ray temperature $8.6_{-1.2}^{+1.3}$ keV of the cluster recently
measured with $Chandra$ is consistent with this high lensing mass.
When we adopt the 1-$\sigma$ lower limit as a mass threshold and use the cosmological parameters
favored by the Wilkinson Microwave Anisotropy Probe 5-year (WMAP5) result,
the expected number of similarly massive clusters at $z\gtrsim1.4$ in the $11$ square degree survey is $N\sim5\times10^{-3}$.
Therefore, the discovery of the cluster within the survey volume is a rare event 
with a probability $\lesssim 1$\%, and may open new scenarios in our current
understanding of cluster formation within the standard cosmological model.
\end{abstract}

\altaffiltext{2}{Department of Physics, University of California, Davis, One Shields Avenue, Davis, CA 95616, USA}
\altaffiltext{3}{European Southern Observatory, Karl-Schwarzschild-Strasse 2, D-85748, Garching, Germany}
\altaffiltext{4}{Department of Physics and Astronomy, Johns Hopkins University, 3400 North Charles Street, Baltimore, MD 21218.}
\altaffiltext{5}{Department of Physics and Astronomy, University of Utah, Salt Lake City, UT 84112, USA}
\altaffiltext{6}{European Southern Observatory, Alonso de Cordova 3107, Casilla 19001, Santiago, Chile}
\altaffiltext{7}{E.O. Lawrence Berkeley National Laboratory, 1 Cyclotron Rd., Berkeley, CA 94720, USA}
\altaffiltext{8}{Department of Astronomy, Universidad de Concepci\'on. Casilla 160-C, Concepci\'on, Chile}
\altaffiltext{9}{National Radio Astronomy Observatory, 1003 Lopezville Road Socorro, NM 87801}
\altaffiltext{10}{Wachovia Corporation, NC6740, 100 N. Main Street, Winston-Salem, NC 27101}
\altaffiltext{11}{Max-Planck-Institut für extraterrestrische Physik (MPE), Giessenbachstrasse 1, 85748 Garching, Germany }


\section{INTRODUCTION \label{section_introduction}}

Despite quite a few concerted efforts over the past decade, 
the number of confirmed X-ray emitting clusters at redshifts beyond unity is still small.
This rarity is not surprising when we consider both the hierarchical build-up of structures
(i.e., evolution of mass function) and the observational challenges (i.e., cosmological
surface brightness dimming).
Because the abundance of these $z>1$ clusters is extremely sensitive to cosmological
parameters (particularly, the matter density of the universe $\Omega_M$ and its fluctuation $\sigma_8$),
every individual cluster in this redshift regime deserves a careful study of 
its observables closely related to the mass properties.

Although still efficient in finding massive $z>1$ clusters, X-ray observations alone do not provide
a secure constraint on their masses. This is not only because the X-ray photons are scarce in this redshift regime,
but also because a significant fraction of the $z>1$ clusters are likely to deviate
from hydrostatic equilibrium: the key justification for the use of X-rays in cluster mass estimation.
Gravitational lensing is thus highly complementary to the X-ray technique in that the method
does not rely on the dynamical state of the mass that it probes.
This advantage over other
methods grows with increasing redshift (while there still remain a sufficient number of background galaxies)
because we expect to find more and more unrelaxed, forming clusters
in the younger and younger universe. One caveat, however, is that the mass estimation using only weak-lensing 
suffers from the so-called mass sheet degeneracy, which effectively induces an additional uncertainty
in the mass determination. Nevertheless, this degeneracy can be lifted by imposing that the cluster mass profile is 
approximated by a parameterized
mass profile, and this assumption is less dangerous than the equilibrium hypothesis in X-ray studies
when non-thermal X-ray emission is likely to play an important role.

However, it is observationally challenging to perform a lensing analysis of $z>1$ clusters.
The reason is two-fold. First, as the lenses are already at high redshifts, the number of background
galaxies is substantially small. Second, the shapes of these background galaxies,
only slightly larger than a typical seeing in ground-based observations, are highly subject to
systematics of the instrument. This is why deep, space-based imaging is essential in weak-lensing analysis
of high-redshift clusters. Until now, only three $z>1$ clusters, 
namely RDCS 1252.9-2927 (Lombardi et al. 2005), Lynx-W, and Lynx-E (Jee et al. 2006),
have been measured through weak-lensing with the Advanced Camera for Surveys (ACS) on board
{\it Hubble Space Telescope}. These studies demonstrate the ability of ACS to
detect lensing signals for $z>1$ objects with moderately deep exposures ($\sim5~\sigma$ at 28 ABmag). 

In this paper, we present a weak-lensing analysis of the high-redshift cluster XMMU J2235.3-2557 (hereafter XMM2235) at $z\simeq1.4$
using deep ACS images. It is the highest-redshift cluster known to date measured with weak-lensing; the
previous record-holder is Lynx-W at $z=1.27$ (Jee et al. 2006).
The cluster was discovered in a serendipitous cluster search in archival XMM-$Newton$
observations [as part of the XMM-Newton Distant Cluster Project (XDCP) survey, Mullis et al. 2005]. 
The high X-ray temperatures of $6.0^{+2.5}_{-1.8}$ keV from XMM-Newton 
and the velocity
dispersion of $765\pm265 ~\mbox{km}~\mbox{s}^{-1}$ from 12 galaxies in the redshift range $1.38<z<1.40$ (Mullis et al. 2005) indicate that
XMM2235 might be a massive cluster at $z=1.4$. The recent $Chandra$ measurement of $T_X=8.6_{-1.2}^{+1.3}~\mbox{ke}~\mbox{V}$ 
(Rosati et al. 2009) suggests an even higher mass. 
As the X-ray morphology is nearly symmetric and there is a presence of a significant cool core (Rosati et al. 2009), 
the high temperature 
might be attributed to a high mass rather than a possible merging activity.
Because the expected number density of such a massive cluster is extremely low at $z=1.4$, 
the mass by itself 
(even though it is a single cluster) can have interesting
implications on cosmological parameters. However, this interpretation 
needs an independent confirmation by weak-lensing, which as mentioned above 
gives a reliable mass estimate without any assumption on the dynamical phase of the cluster when the universe
is one-third of its current age.

Throughout the paper, we use a $(h,\Omega_M,\Omega_{\Lambda})=(0.7,0.3,0.7)$ cosmology unless explicitly stated otherwise.
The plate scale is $\sim8.4$~kpc~$\mbox{arcsec}^{-1}$.
All the quoted uncertainties are at the 1-$\sigma$ ($\sim68$\%) level. All magnitudes are given in the
AB system (Oke \& Gunn 1983).

\section{OBSERVATIONS \label{section_obs}}

\subsection{Basic ACS Data Reduction}
XMM2235 was observed with Wide Field Camera (WFC) both as a part of Guaranteed Time Observation (GTO; PROP ID
10698; Ford 2005) and as a part of Guest Observation (GO; PROP ID 10496; Perlmutter 2005) during the periods of
2005 June$-$July and 2006 April$-$July, respectively. The total exposures in F775W and F850LP
(hereafter $i_{775}$ and $z_{850}$, respectively) are 8,150 s and 14,400 s, respectively.
Low level CCD processing (e.g., overscan, bias, and dark substraction, flat-fielding, etc.) was performed 
with the standard STScI CALACS pipeline (Hack et al. 2003),
utilizing the latest available WFC calibration data. 
On the other hand, the high-level science images
were created through the ``apsis'' pipeline (Blakeslee et al. 2003). The main tasks of apsis
include geometric distortion correction via drizzle (Fruchter \& Hook 2002), sky substraction, 
cosmic ray removal, and precise image
alignments. The precise shift
measurements are important and the current alignment
accuracy ($\lesssim0.02$ pixels) of apsis through the ``match'' program (Richmond 2002) 
meets the weak-lensing requirement. We use the Lanczos3 kernel in drizzling with the native ACS pixel scale 
(0.05$\arcsec/$pixel).
The Lanczos3 kernel closely mimics the sinc interpolation, which is the theoretically optimal
interpolation method. The merits of the kernel include sharper PSFs and less noise correlations.

\begin{figure}
\plotone{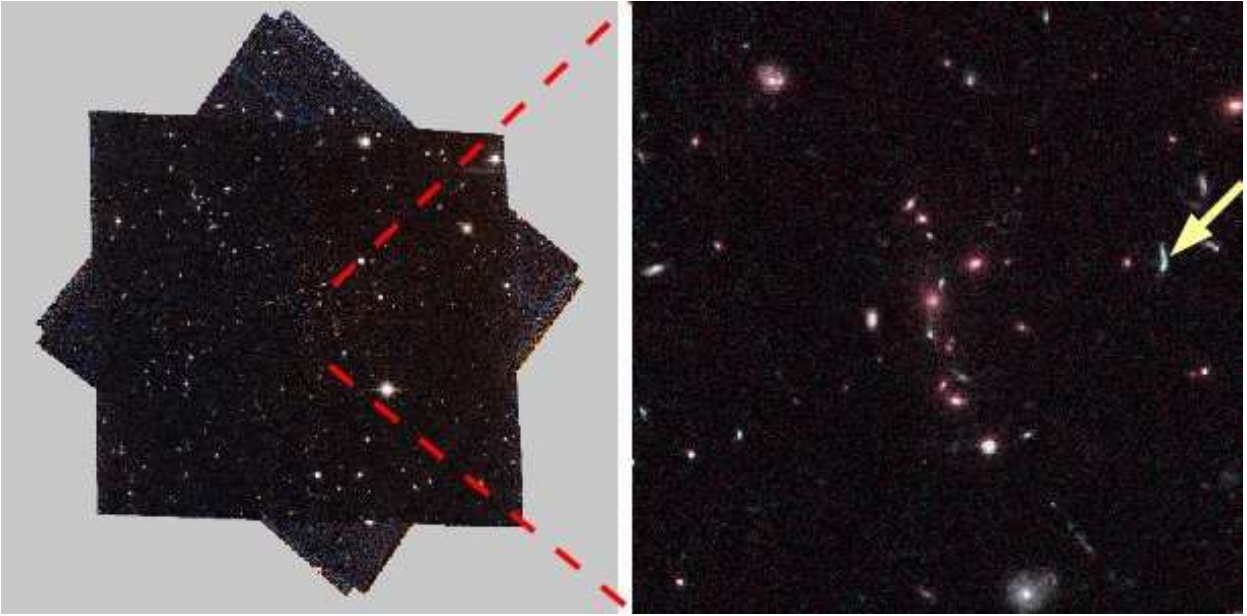}
\caption{HST/ACS color composite of XMM2235 in the observed orientation: North is up and east is left.
We represent the intensities in blue and red with the $i_{775}$ and $z_{850}$ images, respectively.
For the green channel we use the average of the two.
The left panel is to illustrate all the camera orientations used to observe the cluster.
The central $30\arcsec \times 30\arcsec$ region approximately centered on the BCG is shown in detail in the
right panel. The object pointed by the yellow arrow is an arc
candidate (see \textsection\ref{section_mass_estimation} for the detailed discussion of the possibility).
\label{fig_xmm2235}}
\end{figure}

Figure~\ref{fig_xmm2235} shows the pseudo-color composite of XMM2235
created from the pipeline output (north is up and east is left). 
We use the $z_{850}$ and $i_{775}$ images to represent
the red and blue intensities, respectively, while the average of the two filters
is chosen to show the green intensity.
Some exposures in PROP ID 10496 were taken 36\degr$\sim$40\degr~rotated with respect 
to the PROP ID 10698 data, and
therefore the outline of the combined image is roughly an 8-cornered star (left panel).
The right panel displays the blown-up image of the central $30\arcsec \times 30\arcsec$ region approximately
centered on the cluster BCG. We observe the overdensity of red early-type galaxies whose colors are
consistent with that of the red-sequence at $z=1.4$. Also seen are
some indications of strong-lensing features; the yellow arrow points at the arc candidate with 
a tentative redshift of $\simeq3.3$.

\subsection{Object Detection and Photometry}
We performed object detection and photometry by running SExtractor (Bertin \& Arnouts 1996)
in dual image mode. That is, the objects are detected on the same detection image whereas
the photometry is performed on each filter image, which enables us to define
consistent isophotal areas between the $i_{775}$ and $z_{850}$ filters.
The detection image was created by weight-averaging the $i_{775}$ and $z_{850}$ images.
We searched for at least five connected pixels brighter than the sky rms by a factor of 1.5.
After manually discarding $\sim540$ spurious objects (e.g., saturated stars,
diffraction spikes, uncleaned cosmic
rays near field boundaries, H II regions inside nearby galaxies, etc.), 
we obtained a total of 3,035 objects. SExtractor's MAG\_ISO 
was used to calculate colors. For other purposes we adopt 
SExtractor's MAG\_AUTO as the object's total magnitude.

\subsection{PSF Modeling and Shape Measurement}

Weak-lensing extracts signals from subtle distortion of the shapes of background galaxies and thus the success
lies in one's ability to separate intrinsic gravitational lensing distortion from other systematic effects. 
Of the most important systematics is the effect caused by the PSF of the instrument. A PSF is seldom isotropic and its
ellipticity, if uncorrected, induces a false lensing signal. In addition, even if the PSF is isotropic, it dilutes
the ellipticites of source galaxies, which should be taken into account for quantitatively correct interpretation
of the lensing signal.
Although ACS PSFs are smaller than any of existing ground-based telescopes, it is still important to correctly
model the PSF to make the most of the superb resolution of the ACS. The level of the PSF correction determines
the size of the smallest galaxies (thus the number density) that can be used for a lensing analysis.

It has
been noted that ACS PSFs vary across the field and that also this position-dependent pattern changes over time (Krist
2003). Fortunately, the PSF pattern seems to be repeatable (Jee et al. 2007b). From extensive studies
of more than 400 stellar field images, we found that when two observations were taken under a nearly identical
condition (we believe that the focus is the dominant factor, but other effects such as tilting
of the detector plane might be also present), their PSF patterns resemble each other closely.
This repeatability enables us to sample ACS PSFs from dense stellar fields and apply them to weak-lensing fields, where
high S/N stars are sparse (e.g., Jee et al. 2005a; 2005b; 2006; 2007a; see also Schrabback et al. 2007 for an independent analysis
and application to cosmic shear data). 

In this paper, we use the ACS PSF model of Jee et al. (2007b). The model uses the basis functions obtained from a principal component
analysis (PCA) whereas our older PSF model (Jee et al. 2005a; 2005b; 2006; 2007a) relied on shapelets 
(Bernstein \& Jarvis 2002; Refregier 2003).
Although shapelets are reasonably efficient in orthogonal expansion of ACS PSFs, PCA by construction provides the
most compact basis functions, which captures more details with a 	much smaller number of basis functions. This PSF model
based on the PCA was recently applied to our weak-lensing analysis of CL J1226+3332 at $z=0.89$ (Jee et al. 2008).

\begin{figure}
\plotone{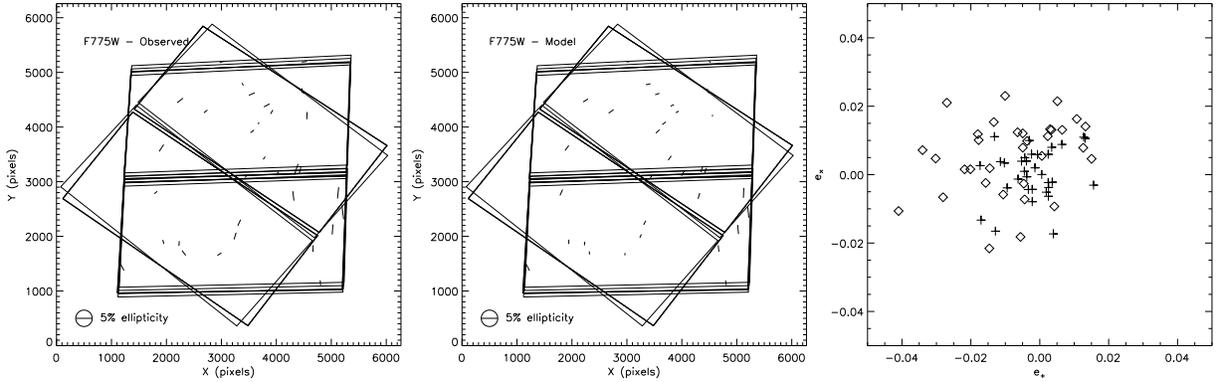}
\caption{Comparison of the observed PSF ellipticities (left) with the predicted
PSF values (middle) from the model for the $i_{775}$ image (unlike in Figure~\ref{fig_xmm2235},
we use the observed orientation here, which is a preferred choice in a PSF analysis; north is right and east is up).
The comparison shows that our final PSF model, the result of stacking different
PSF models for individual exposures, is robust and very close to the observed PSF pattern in the final
dithered image.
The plot on the right-hand side shows the ellipticity components ($e_{+},e_{\times}$) of the observed
PSFs (diamond) and the residuals (`+' symbol) calculated by subtracting the model PSFs from the
observed PSFs.
The mean deviation between the model and the observations is $< |\delta e| >=0.009\pm0.006$.
The center of the residual distribution is $(\delta e_1,\delta e_2)=(-2\times10^{-3},-1\times10^{-3})$.
\label{fig_psf_matching}}
\end{figure}

Our stacked images of the $i_{775}$ and $z_{850}$ filters consist of 10 and 35 exposures, respectively. Because we choose
to measure object shapes in the final stacked image (alternatively, one can estimate a shape exposure-by-exposure
and later combine all the shapes), we need to model the PSFs in a somewhat complicated way. 
The PSF of a given object in the final science image is the result of superposition of many PSFs from different exposures.
Hence, we first determined a PSF pattern for each exposure by finding a best-matching PSF template
from the PSF library by comparing the shapes of the high S/N stars to the model.
To account for charge transfer inefficiency effect, the PSFs in this template are slightly modified at this stage 
(see \textsection\ref{section_cti_correction} for details of our CTI correction method). 
Then, we applied the required offsets and 
rotations to this set of PSFs in a similar way that we were registering individual exposures to the final mosaic.
Finally, we combined 
all the contributing PSFs by weighting each PSF with the exposure time.
Comparison of these model PSFs with the observed PSFs (star images in the stacked images) provide an important
sanity check. Figure~\ref{fig_psf_matching} shows that our $i_{775}$ PSFs constructed in this way (left) closely represent the observed
PSFs (middle).  
The residual ellipticities shown on the right panel are small $<|\delta \epsilon|^2>^{0.5}\simeq0.01$
and have no preferential direction. 

We found that the shape analysis result in the $z_{850}$ image is unsatisfactory, and we decided to confine our shape analysis to
the $i_{775}$ image. This is because in the $z_{850}$ image (1) the PSFs are larger (thus giving poorer constraint for small objects),
(2) the shapes are more color-dependent (note that our PSF model for the $z_{850}$ filter represents a mean color of the used globular cluster), 
and (3) the source population that we choose to use has substantially lower S/N value in the $z_{850}$ passband 
despite the $\sim80$\% longer integration time.
In principle, one can try to combine the shapes from both 
the $i_{775}$ and $z_{850}$ images by assigning properly higher weights to the $i_{775}$ shapes. However, it is non-trivial
to rigorously model all the contributing errors (especially, color-dependent PSF shapes). 

We measure galaxy ellipticities by fitting a PSF-convolved elliptical Gaussian
to the images. Mathematically, this is equivalent to the method of Bernstein \& Jarvis (2002), who proposed to iteratively
shear objects to match a $circular$ Gaussian until the quadrupole of the object vanishes.
They implemented the method first by decomposing galaxy shapes with shapelets and then by applying
shear operators to the shapelet coefficients until the object becomes round. We noted in Jee et al. (2007a),
however, that directly fitting an elliptical Gaussian to the pixelized object reduces aliasing compared
to the shapelet formalism, particularly when the object has extended features.
Also, fitting a PSF-convolved elliptical Gaussian to pixelated images
is more numerically stable for faint objects, and provides straight-forward error estimates in the resulting ellipticity.

\subsection{Charge Transfer Inefficiency Correction \label{section_cti_correction}}

After charges are collected in each pixel of a CCD, they are transferred pixel-to-pixel	in the column direction 
to the readout register.
During the transfer, charge losses and redistribution occur because defects in the silicon undesirably trap 
and then release with a fast time constant some fraction
of the charges. The effect is called charge transfer inefficiency\footnote{Quantitatively,
charge transfer inefficiency is defined as the fraction of charge left behind in a single pixel transfer.} (CTI) 
and is an important parameter 
in the characterization of the CCD performance. As the defects result mostly from radiation damage, CTI increases
with time. In addition, because charge trapping happens every time charges are transferred across a trap,
CTI is greater for pixels farther from the readout register. 
The release of the trapped charges during readout
(with a short time constant that depends on the CCD temperature)
redistributes the trapped charges into a "tail" along
the column that points away from the shift direction.
When there is significant radiation induced charge trapping, 
this cascading effect is visually identified as CTI trails.
In aperture photometry, CTI leads to underestimation of source fluxes because obviously some charges trail outside the aperture.
In weak-lensing analysis, where stringent shape measurement is required in addition to careful photometry,
this charge trailing, if found significant, poses an additional difficulty in systematics control.

Riess and Mack (2004) reported that strong evidence for photometric losses is found in the parallel direction for ACS
observations. As is the case for a typical CTI effect, they observed that ACS CTI grows 
with decreasing stellar flux and background, as well as with time. As a result, their CTI characterization
includes these parameters and the prescription is useful for aperture photometry.
Rhodes et al. (2007) discussed the ACS CTI issue in the context of weak-lensing and derived
an empirical prescription to correct object ellipticity for the CTI-induced elongation. 

In the current work, we independently investigate the CTI-induced charge trailing effect of ACS in object
ellipticity measurement. Although the simple form of the solution by Rhodes et al. (2007) is appealing, their
prescription cannot be readily applied to our analysis because their shape measurement and PSF correction
schemes are different from ours. 

Although it is possible to study the CTI-induced elongation from astronomical objects by assuming that without
CTI the mean ellipticity over a large number of objects must vanish, we measure CTI-induced shape distortion using
cosmic-rays in the FLT (pre-drizzled) images of XMM2235. Because cosmic-rays are not affected by the instrument PSF,
the systematic elongation of the cosmic-rays due to the CTI in the FLT images is 
nicely disentangled from the effect of the imperfection in PSF modeling
and the residual geometric distortion. In addition, cosmic-rays are numerous, and thus we can obtain useful statistics
directly from the XMM2235 images themselves, which obviates the need for external calibration.

Cosmic-rays were
selected by looking for objects whose half-light radii are less than the stars and
fluxes are greater 200 electrons; the flux distribution of cosmic rays are independent of exposure time and has
a well defined lower limit.
This selection method, of course, misses
a number of cosmic rays that occupy multiple bright pixels. However, this scheme still gives us a sufficient number of cosmic-rays
for a statistical analysis.
We quantify the CTI-induced elongation with the mean ellipticity of cosmic-rays. But for
any systematics, the mean ellipticity should be consistent with zero; we
used unweighted moments to compute ellipticity $[(a-b)/(a+b)]$ of cosmic-rays.

\begin{figure}
\plotone{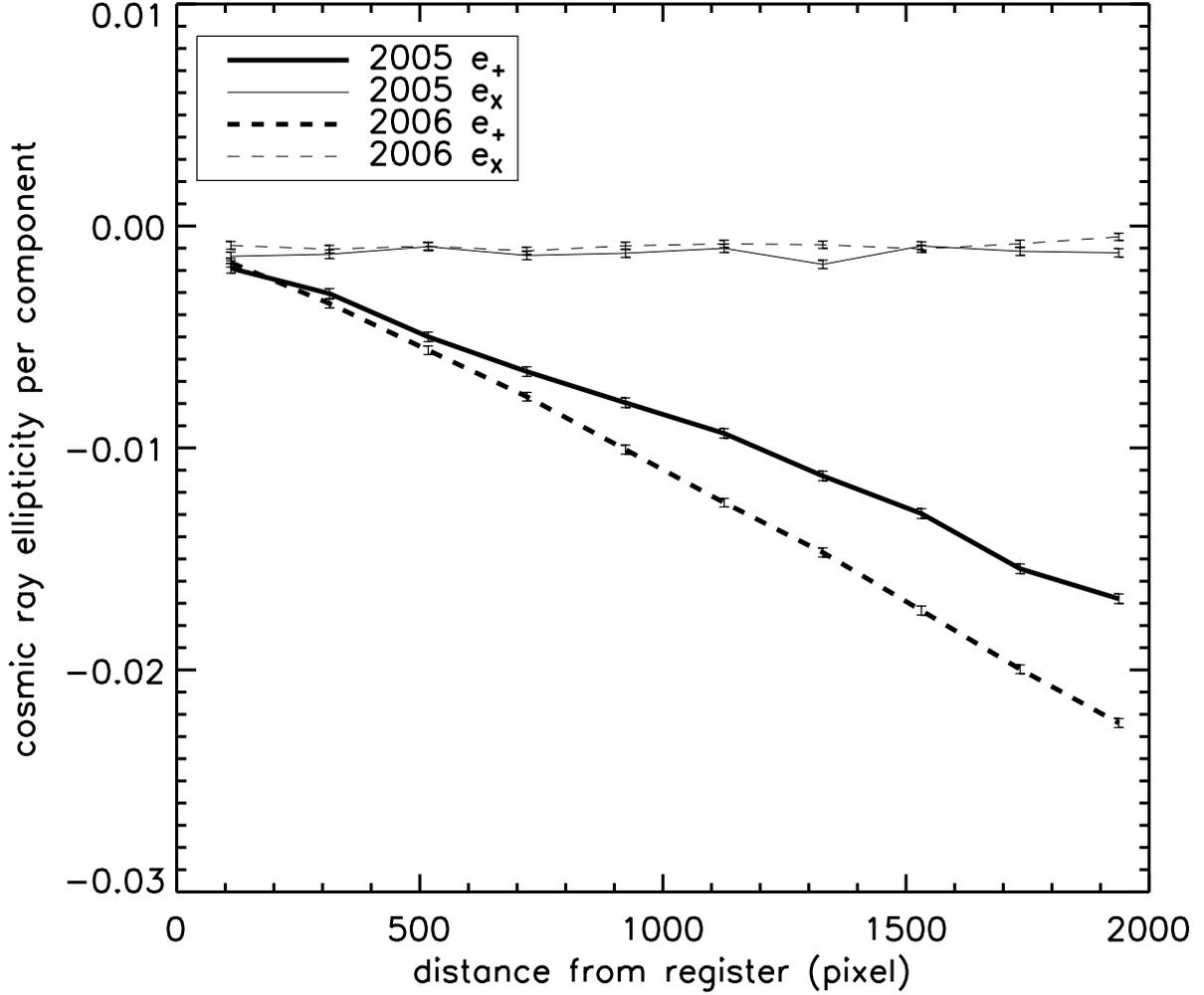}
\caption{Elongation of cosmic rays by CTI in WFC CCDs. The XMM2235 data were taken in two periods of time and here we
show the CTI-induced elongation measured separately for each period. 
The symbols $e_+$ and $e_{\times}$ represent the ellipticity in the horizontal (vertical if negative) and
the $45$\degr~(135$\degr$ if negative) directions, respectively.
The year 2005 and year 2006 datasets
contain $\sim280,000$ and $\sim670,000$ cosmic rays, respectively. We measured ellipticity using
unweighted moments in the FLT files and the WFC1 and WFC2 results were combined.
As expected, the CTI-induced elongation ($e_+$) linearly depends on
the number of charge transfers (distance from readout registers) and the time elapsed since the installation of
ACS. The $e_{\times}$ component does not change as a function of the number of charge transfers, remaining close to zero. However, there
are sub-percent ($\lesssim0.002$) level biases toward negative $e_{\times}$ (elongation in the 135 $\degr$ direction), whose
origins are unclear at this moment.
\label{fig_cte_vs_time}}
\end{figure}

Figure~\ref{fig_cte_vs_time} shows the average $e_{+}$  and $e_{\times}$ components of the ellipticity of cosmic-rays whose S/N is greater than 10
as a function of distance
from the read-out register for two data sets. The first dataset ($solid$) containing $\sim280,000$ cosmic-rays is the XMM2235 data taken
during the 2005 June-July period and the second dataset (dashed) containing $\sim670,000$ cosmic-rays 
was taken during the 2006 April-July period.
The linear dependence of the charge elongation on the distance from the readout register is clearly observed (thick line).
In addition, the slope is steeper for the second dataset, which is taken approximately 1 year after
the first dataset. When we assume that the slope is zero at the installation of ACS in the year 2002, it seems that
the CTI slope increases almost linearly with time. These linear dependence of the CTI-induced elongation on
both time and distance from register is consistent with the result of Rhodes et al. (2007).
The $e_{\times}$ component (thin line) does not change as a function of the number of charge transfers as expected. However, there
are sub-percent ($\lesssim0.002$) level biases toward the negative value (elongation in the 135 $\degr$ direction with respect to the x-axis), whose
origins are unclear.

\begin{figure}
\plottwo{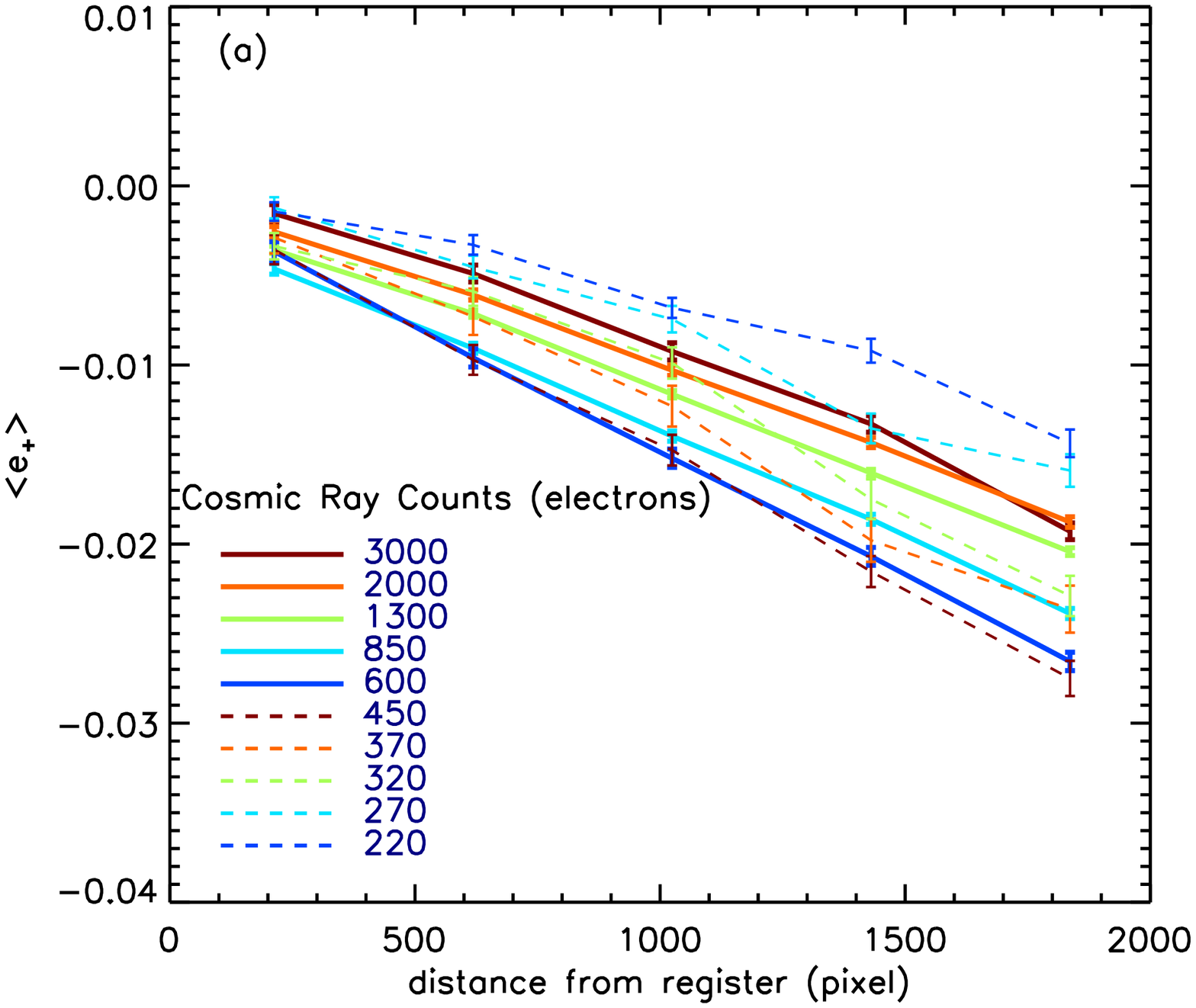}{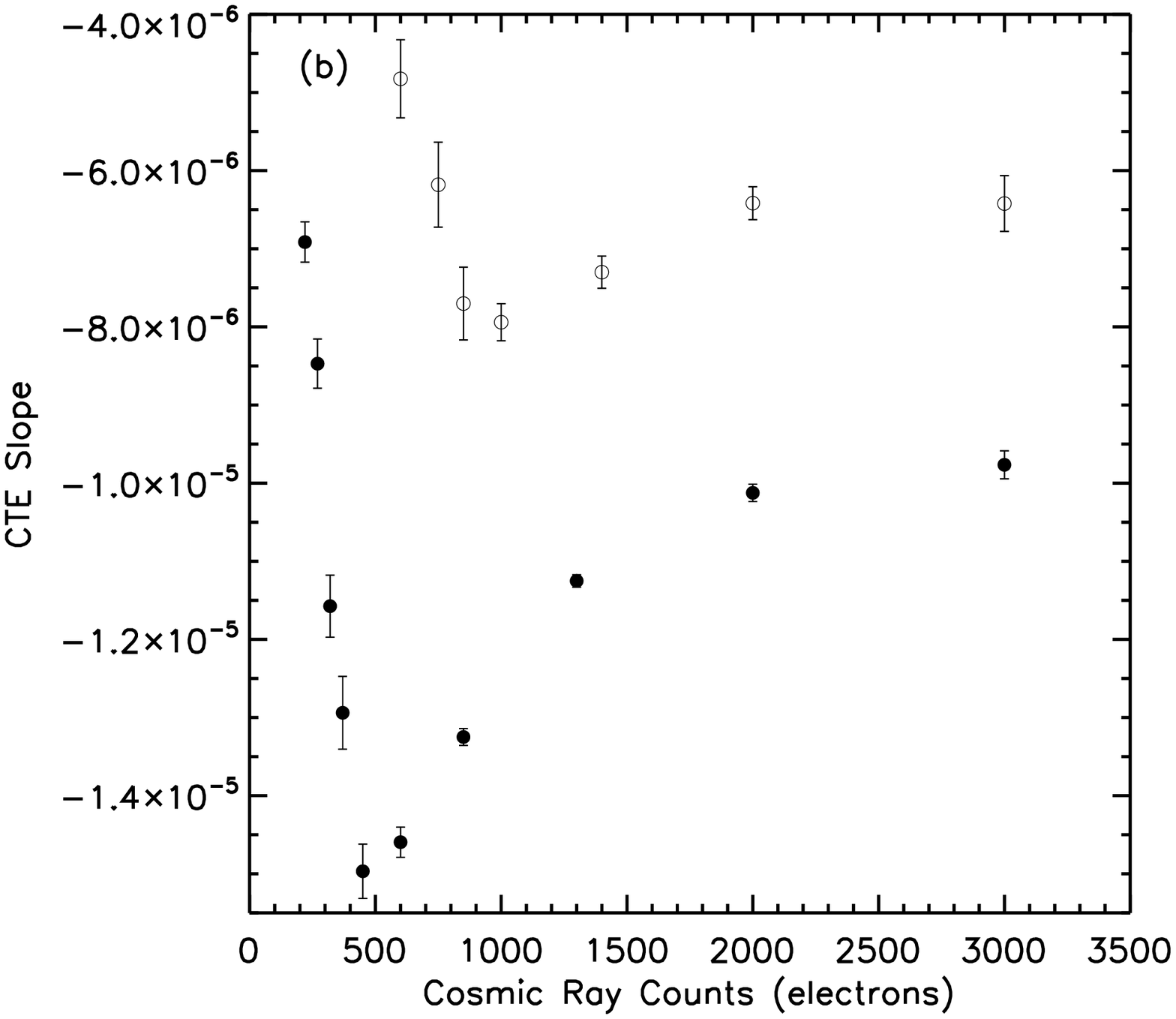}
\caption{Flux dependence of CTI-induced elongation. (a) We divided the cosmic-rays 
of the year 2006 dataset into 10 groups according to their flux (counts) and 
measured their elongation as a function of distance from the readout register.
For counts greater than $\gtrsim 500$, the slope steepens for decreasing flux.
However, the trend is reversed for counts $\lesssim 500$.
(b) The slope of the CTI-induced elongation versus cosmic ray counts.
Closed circles show the slopes of the curves in (a) against cosmic ray counts; the more
negative the severer the CTI effect. It is easier here to observe the flux dependence of
the CTI effect. We found that the count at the turnaround 
is determined by the background level. Open circles are for the subset of the 2005 data, which
has longer exposures; the mean background here is $\sim80$ electrons whereas
the mean background in the 2006 dataset is $\sim20$ electrons.
The turnaround happens at $\sim1000$ electrons.
\label{fig_cte_vs_snr}}
\end{figure}

Now to investigate the flux dependence we divided the cosmic-rays into 10 groups according to their flux (counts) and
repeated the above analysis.
In Figure~\ref{fig_cte_vs_snr}a, we display the result for the year 2006 dataset. Note that for counts greater than $\gtrsim500$ 
the slope steepens for decreasing object counts. However, for counts$\lesssim500$, the trend is reversed.
It is easier here to observe the flux dependence of the CTI effect in Figure~\ref{fig_cte_vs_snr}b, where we
plot the slope of the CTI-induced elongation against cosmic ray counts.
Closed circles show the slopes of the curves in Figure~\ref{fig_cte_vs_snr}a versus cosmic ray counts;
the more negative the severer the CTI effect. It is clear that the flux dependence changes its trend
at counts $\sim500$. This threshold is found to depend on the background level, shifting toward
higher values for higher backgrounds. As an example, we also display the measurement (open circle) for the subset 
of the 2005 data, which has longer exposures; the mean background here is $\sim80$ electrons whereas
the mean background in the 2006 dataset is $\sim20$ electrons. The reverse of the trend happens at $\sim1000$ electrons in this case.
Although not shown here, from the archival ACS data we also noticed that for exposures as short as $\sim30$s the turnaround
is not observed at least for counts $\gtrsim$200 (because of the lower limit of the cosmic ray flux, we could not
probe the $\lesssim$200 regime).
 
For a sanity check we investigated also the $e_{\times}$ components of the ellipticity of the cosmic rays. If our
selection criterion (half light radius) creates a bias toward low ellipticity for low flux cosmic rays, this can
falsely masquerade as the mitigation of the CTI. Our analysis, however, shows that no gradual selection bias (as the S/N value decreases) 
toward low ellipticity is found in the above flux range.
 
Through photometry Riess \& Mack (2004) reported a CTI mitigation for faint stars ($\lesssim$1000 electrons) in
the presence of substantial background ($\gtrsim$a few tens of electrons/pixel).
Nevertheless, their CTE parameterization
($\propto~SKY^B\times FLUX^C$) with one set of parameters does not seem to indicate the
presence of the explicit turnaround that we report here. However, our independent
analysis of their datasets suggests that the effect might be also photometrically observable although
the signal is weaker than in the current cosmic-ray analysis. 
We refer readers to \textsection\ref{section_appendix_cti} for the details of our independent test results.

Our finding of this non-monotonic dependence of the CTI-induced elongation on object counts is also different
from the Rhodes et al. (2007)'s empirical correction, which assumes that the CTE-induced elongation increases continuously
as the inverse of the objects' S/N value.
In our shape analysis, instead of attempting to derive an analytic solution, we arrange the above results in tables
and use them by interpolation for each object.

We choose to account for the CTI trailing effect by applying the correction to
PSFs rather than to object ellipticity. 
Because we measure object shapes with the stacked PSF that result from different
exposures at different orientation, this CTI correction scheme can be incorporated into our 
existing PSF correction method.
Although the CTI-induced elongation
is not strict convolution, we approximate the effect by stretching the PSF in the
y-direction in the
CCD coordinate.
We derived the stretching factor numerically through image simulations using real galaxy images.

Although our quantification of the CTI effect above is the result of our time-consuming efforts,
we clarify here that,
because the effect is small, the cluster weak-lensing analysis result presented hereafter is not significantly
altered even if we omit the correction. 
As is detailed in \textsection\ref{section_source_selection}, we select faint blue galaxies
[$(i_{775}-z_{850})<0.5$ and $24<z_{850}<29$] as the source population. The mean
S/N of these objects is $\sim10$ in the stacked $i_{775}$ image (therefore, their S/N
in individual exposure is much lower). According to the current analysis,
we estimate that the ellipticity of the galaxies farthest from the register needs to be adjusted by $\sim0.01$ on average. 
Of course, this effect is small and thus negligible for common cluster weak-lensing.

However, we stress that when one is looking for
a much weaker signal on large scales (e.g., cosmic shear or galaxy-galaxy lensing) all these issues
discussed above become critical in the quantification of the lensing signal, as well as in the removal of the B-mode signal.

\section{WEAK-LENSING ANALYSIS}

\subsection{Basic Weak-Lensing Theory \label{section_theory}}

Many excellent reviews on the topic are available in the literature and here we only summarize the 
basic weak-lensing theory suitable for the subsequent description. In a weak-lensing regime, where
the characteristic length of the lensing signal variation is larger than the object size, the
shape distortion is linearized as follows:
\begin{equation}
\textbf{A}(\bvec{x}) =  \delta_{ij} - \frac {\partial^2 \Psi (\bvec{x})} {\partial x_i \partial x_j}
= \left ( \begin{array} {c c} 1 - \kappa - \gamma _1 & -\gamma _2 \\
                      -\gamma _2 & 1- \kappa + \gamma _1  
          \end{array}   \right ),  \label{eqn_lens}
\end{equation}
\noindent
where $\textbf{A}(\bvec{x})$ is the transformation matrix $\bvec{x}^\prime = \textbf{A} \bvec{x}$, which
relates a position $\bvec{x}$ in the source plane to a position $\bvec{x}^\prime$ 
in the image plane,
and $\Psi$ is the 2-dimensional lensing potential. 
The convergence $\kappa$ is the surface mass density in units of critical surface mass density
 $\Sigma_c=c^2 
D(z_s)/ (4\pi G D(z_l) D(z_l,z_s))$,
where $D(z_s)$, $D(z_l)$, and $D(z_l,z_s)$ are the angular diameter distance from the observer to the source,
from the observer to the lens, and from the lens to the source, respectively.
The convergence $\kappa$ and the shears $\gamma_{1(2)}$ are related to the lensing potential $\Psi$ via
\begin{eqnarray}
\kappa=\frac{1}{2}  (\psi_{11}+\psi_{22}) \label{eqn_kappa},~
\gamma_1=\frac{1}{2} (\psi_{11} -\psi_{22}) \label{eqn_shear1},~
\mbox{and}~
\gamma_2=\psi_{12}=\psi_{21} \label{eqn_kappa_shear},
\end{eqnarray}
\noindent
where the subscripts on $\psi_{i(j)}$ denote partial differentiation with respect to $x_{i(j)}$.

Equation~\ref{eqn_lens} implies that a circular
object gains an ellipticity\footnote{The ellipticity is defined as (a-b)/(a+b), where $a$ and $b$ are
the major- and minor-axes, respectively}
of $g=\gamma/(1-\kappa)$ under the transformation\footnote{This is valid only if $g<1$ (i.e., in the weak-lensing regime).}. 
Because galaxies come with different shapes and
radial profiles, a rigorous calibration must be made in order to make the practical shape
measurements and the theoretical relation agree.

To a first order, an average ellipticity over a sufficient number of galaxies in a region
samples the underlying shear $g$. If we further assume $g\simeq\gamma$
(this assumption becomes increasingly invalid near the cluster core where $\kappa$ is non-negligible), the surface mass density
$\kappa$ is directly obtained through the following convolution:
\begin{equation}
\kappa (\bvec{x}) = \frac{1}{\pi} \int D^*(\bvec{x}-\bvec{x}^\prime) \gamma (\bvec{x}^\prime) d^2 \bvec{x} \label{k_of_gamma}. 
\end{equation}
\noindent
where $D^*(\bvec{x} )$ is the complex conjugate of the convolution kernel $D(\bvec{x} ) = - 1/ (x_1 - \bvec{i} x_2 )^2$. 

A careful examination of Equation~\ref{eqn_lens} soon reveals, however, that the ellipticity change induced by lensing
is invariant under the $(\kappa \rightarrow 1-\lambda + \lambda \kappa)$ transformation, where $\lambda$ is an arbitrary
constant. This invariance is often called a mass-sheet degeneracy, although the term is misleading as it can 
potentially and inadequately imply that
the lensing signal is invariant under the introduction of a constant mass-sheet when in fact a rescaling of $\kappa$
is also required. In order to break the degeneracy, one must either 1) incorporate additional strong-lensing
data (e.g., more than one multiple image systems  at different redshifts) or 2) assume
a specific $\kappa$ value in some part of the field. In typical weak-lensing analyses, the
second method is often the only available option, and the local $\kappa$ value is determined by fitting
parameterized models to the cluster's reduced tangential shear profile.

Reduced tangential shears are defined as
\begin{equation}
 g_T  = \left< -  g_1 \cos 2\phi - g_2 \sin 2\phi \right > \label{tan_shear},
\end{equation}
\noindent
where $\phi$ is the position angle of the object with respect to the cluster center.
If no shear is present, the average of the tangential shear must vanish (or oscillate
around zero). For a simple axisymmetric halo, we expect the value to reach its maximum
near the Einstein radius and decrease for increasing radius.
Inside the Einstein radius, in principle the signal decreases rapidly, reaches 
its minimum (negative) at the radial critical curve, and vanishes. However, observationally
it is challenging to measure the tangential shears reliably in this regime because of the
small number of the source population, the high cluster member contamination,
the halo substructures, etc.

Alternatively, instead of using the above convergence map, some authors choose to
use the so-called aperture mass densitometry first suggested by Fahlman et al. (1994) and
modified by Clowe et al. (1998).
This method is preferred if the two-dimensional mass reconstruction significantly 
suffers 
from the artifacts mentioned
above. In Jee et al. (2005a; 2005b), where this is not a concern, we demonstrated that this aperture densitometry gives a 
consistent result with
the one from the direct use of the convergence map. Because this consistency is also observed in the current analysis,
we omit the aperture densitometry result in favor of the direct use of the mass map.

\subsection{Source Selection and Redshift Distribution Estimation \label{section_source_selection}}

\begin{figure}
\plottwo{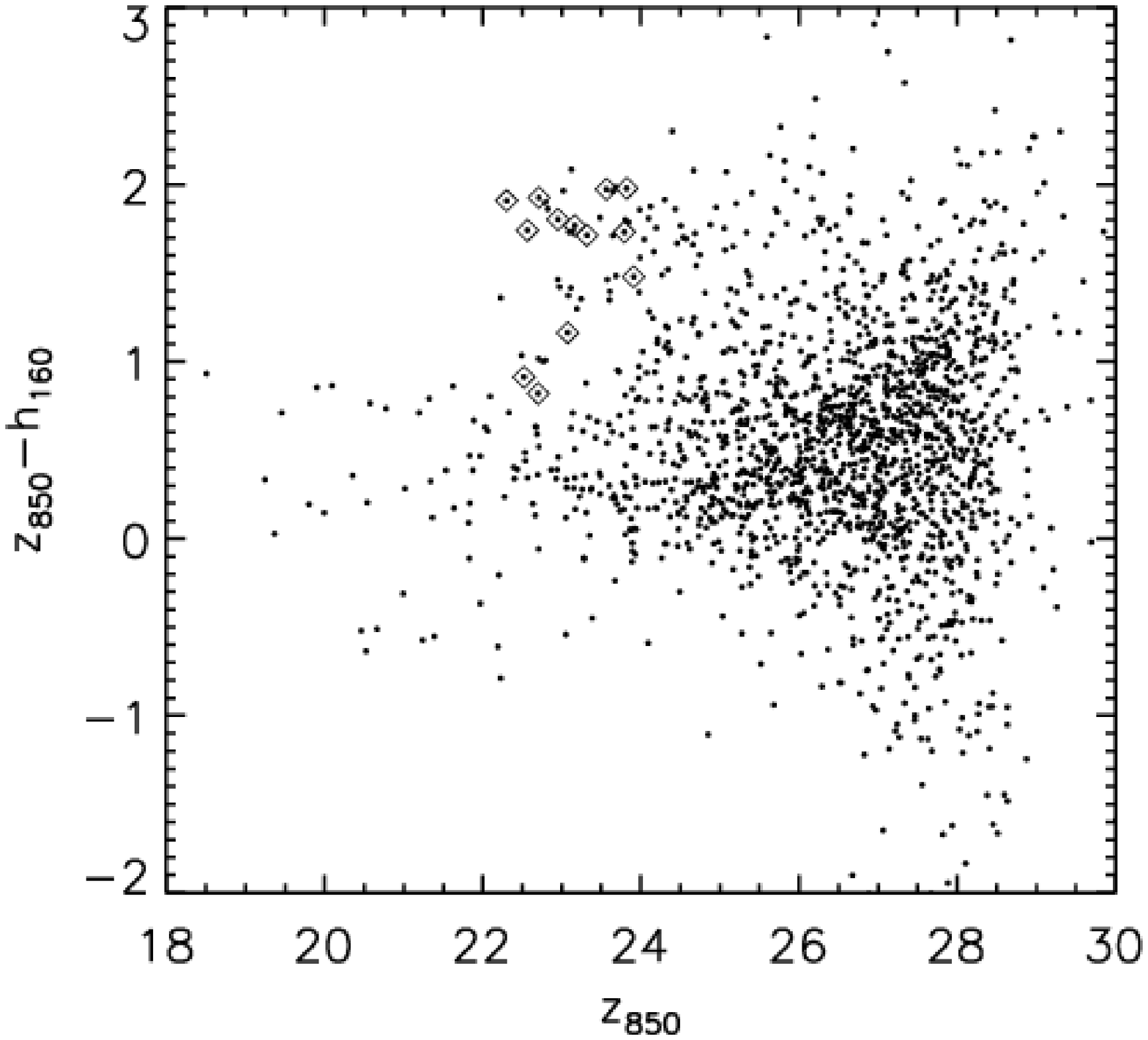}{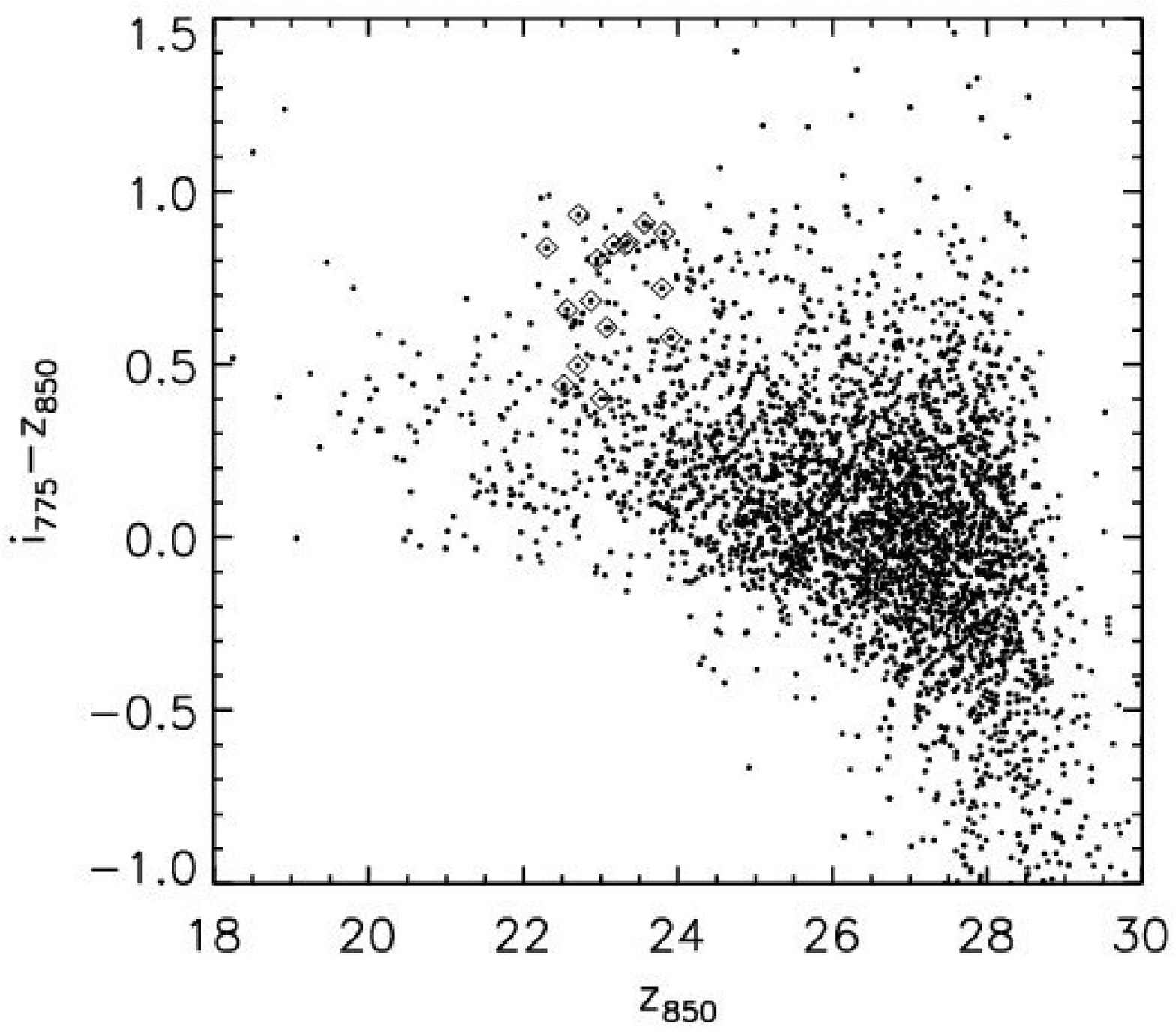}
\caption{Color magnitude relation in the XMM2235 field. The $z_{850}-h_{160}$ color (left)
is a better discriminator of the cluster red-sequence than the $i_{775}-z_{850}$ color (right) at $z=1.4$.
The NICMOS image covers only part of the ACS image and thus fewer
data points are seen in the left panel. 
The diamond symbols represent the spectroscopically
confirmed cluster members ($1.38<z<1.40$).
\label{fig_cmr}}
\end{figure}

With only the two ($i_{775}$ and $z_{850}$) passband ACS images available, we follow the conventional approach:
selection of faint galaxies bluer than the cluster red-sequence. The redshifted 
4000 \AA~break at $z=1.4$ is located
on the red tail of the $z_{850}$ filter transmission curve, and 
hence the $i_{775}-z_{850}$ color is not an optimal
filter combination. 
Nevertheless, the cluster red-sequence of XMM2235 is still visible in the color-magnitude
diagram as among the brightest and reddest (Figure~\ref{fig_cmr}). 
The diamond symbols represent the spectroscopically confirmed
cluster members ($1.38<z<1.40$). We selected source galaxies with $(i_{775}-z_{850})<0.5$ and 
$24<z_{850}<29$ while
discarding objects whose S/N in the $i_{775}$ filter is less than 5.
After further removing objects whose ellipticity uncertainty is greater than $\delta \epsilon > 0.2$, we obtained a total 
of 1554 galaxies 
($\sim120~\mbox{arcmin}^{-2}$). 

\begin{figure}
\plotone{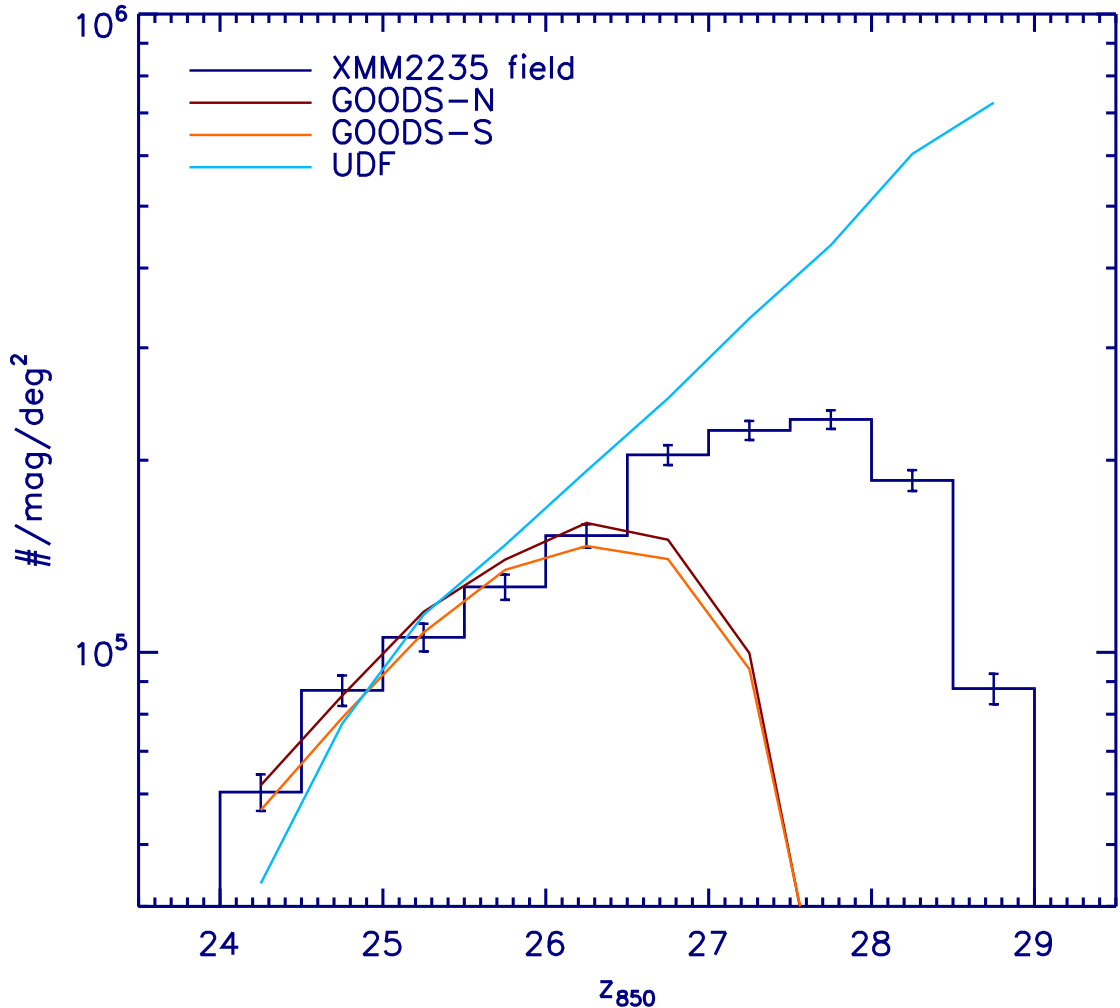}
\caption{Magnitude distribution of source population. We define the source population as galaxies satisfying
the $(i_{775}-z_{850})<0.5$, $24<z_{850}<29$, and $S/N>5$ criteria. Also displayed are the magnitude distribution
of the galaxies in the GOODS North, South, and UDF that are selected by applying the same color and magnitude cut.
The comparison indicates that our source catalog is not likely to be severely contaminated by blue
cluster members of XMM2235.
\label{fig_source_population}}
\end{figure}

A potential problem in this approach is the possible contamination of the source catalog from the blue cluster galaxies.
Considering the so-called Butcher-Oemler effect (Butcher \& Oemler 1984), we expect a large 
fraction of the cluster members to be bluer than the
red-sequence. If the contamination is found to be significant, we must take into account the dilution in the
estimation of the source redshift distribution. However, our test shows that the contamination, if any, is
negligible. In Figure~\ref{fig_source_population}, the solid line shows the normalized number density of objects
in the source catalog. When this is compared with the Great Observatories Origins Deep Survey (GOODS; Giavalisco et al. 2004) ACS data,
no measurable excess is found; we applied the same color and magnitude selection to the objects in the GOODS field.
Because the GOODS images are shallower than our cluster field, the number density turns around at brighter
magnitude ($z_{850}\sim26.5$). For galaxies fainter than this, we used the Ultra Deep Field (UDF; Beckwith et al. 2003) data to
compare the number density. Although the depth difference causes the discrepancy at $z_{850}>26.5$, we
do not observe any excess in the $24<z_{850}<29$ regime, either. This result is also presented
in our previous papers (e.g., Jee et al. 2005a; 2008).
Consequently, we disregard
the dilution effect of the lensing signal by the cluster members (however, remember that the foreground contamination is still
substantial). 

We estimate the redshift distribution of the source population utilizing the publicly available
UDF photometric redshift catalog (Coe et al. 2006). The ultra deep images in six filters
from F435W to F160W provides unprecedentedly high-quality photometric redshift information well
beyond the limiting magnitude of the XMM2235 images.
We binned our entire source population
into 0.5 magnitude intervals and determined the redshift of galaxies in each bin using
the UDF photo-$z$ catalog while taking into account the difference in the number density.
The photometric redshift distribution of source population is often expressed in
terms of $\beta$:
\begin{equation}
\beta = \mbox{max} \left [ 0, \frac{D(z_l,z_s)} {D(z_s)} \right ]. \label{eqn_beta}
\end{equation}
\noindent

We obtain $<\beta>=0.16$ for the entire source population in the
adopted cosmology; the foreground population is estimated to comprises about 45\% of our source galaxies, and
thus the contamination significantly dilutes the lensing signal.
The value $\beta=0.16$ corresponds
to a single source plane at $z_{eff}=1.84$ and gives a critical lensing density of $\Sigma_c = 5950 \mbox{M}_{\sun} \mbox{pc}^{-2}$
at $z=1.4$. 
Because the lens is at a high redshift, the uncertainty in the estimation of the effective redshift of the background
galaxies is an important factor in the error budget; for example, a value of $\delta z_{eff}=0.1$ would introduce a $\sim20$\% uncertainty
in the mass estimation at $z=1.4$ whereas the same $\delta z_{eff}$ value would give rise to only a $\sim1$\% mass error for a $z=0.2$ cluster.
There are three critical issues in the estimation of the $z_{eff}$ uncertainty: 1) the cosmic variance, 2) the resampling error, and 3) 
the difference among the photo-z estimation codes.

In order to assess the first issue, we repeated the above with the three photometric
redshift catalogs obtained from the Hubble Deep Field North (HDF-N), the Hubble Deep Field South (HDF-S), and the Ultra
Deep Field Parallel Field (UDF-P). We used the photometric catalogs of Fernandez-soto et al. (1999) and Labbe et al. (2003) for HDF-N
and -S, respectively, to generate the BPZ catalogs. For UDF-P, we created the catalog using the archival ACS images of the field 
(Blakeslee et al. 2004). 
The resulting scatter for the source population in the XMM2235 field 
is $\delta z_{eff}\simeq0.06$, which translates into a $\sim10$\% uncertainty in the mass estimation. 

Even if the reference field is assumed to be free from the cosmic variance, the application of the photo-z catalog of the reference field to the current 
cluster field inevitably introduces resampling errors. This scatter, however, is relatively small and estimated
from bootstrapping to be $\sim0.03$ (corresponding to a $\sim4$\% error in mass). 

Apart from these cosmic variance and resampling
errors, in general the choice of the estimation method (including the choice of spectral templates and priors) is also important. 
For example, in HDF-N we observe an offset as much as $\delta z_{eff}\sim0.1$ between the BPZ and the 
Fernandez-soto et al. (1999) results (see also Lombardi et al. 2005 for their discussion). However, the difference in results due to the employed method
is substantially mitigated when photometric errors are small. Hence, for the galaxies in the UDF (where the typical $10~\sigma$ limiting magnitude
is $\sim29$), we believe that this issue is not critical. Our experiments show that 
the amplitude of this bias is similar to the resampling error ($\sim0.03$).

Because the three errors above are independent, the total error is given as the sum in quadrature. Therefore, we estimate that about $11$\% error
in mass is introduced by the uncertainty in $z_{eff}$.

For a high-redshift cluster, understanding the width of the redshift distribution is also important in the translation of the observed
signal. We obtain a value of $<\beta^2>=0.056$, which relates observed reduced shears $g\prime$ to true reduced 
shears $g$ via $g\prime = [1 + (<\beta^2>/<\beta>^2-1) \kappa ]g \simeq (1+1.17 \kappa) g$. 
This correction is increasingly important for higher redshifts and we include this relation in the subsequent analysis.

\subsection{Two-Dimensional Mass Reconstruction \label{section_mass_reconstruction}}

\begin{figure}
\plotone{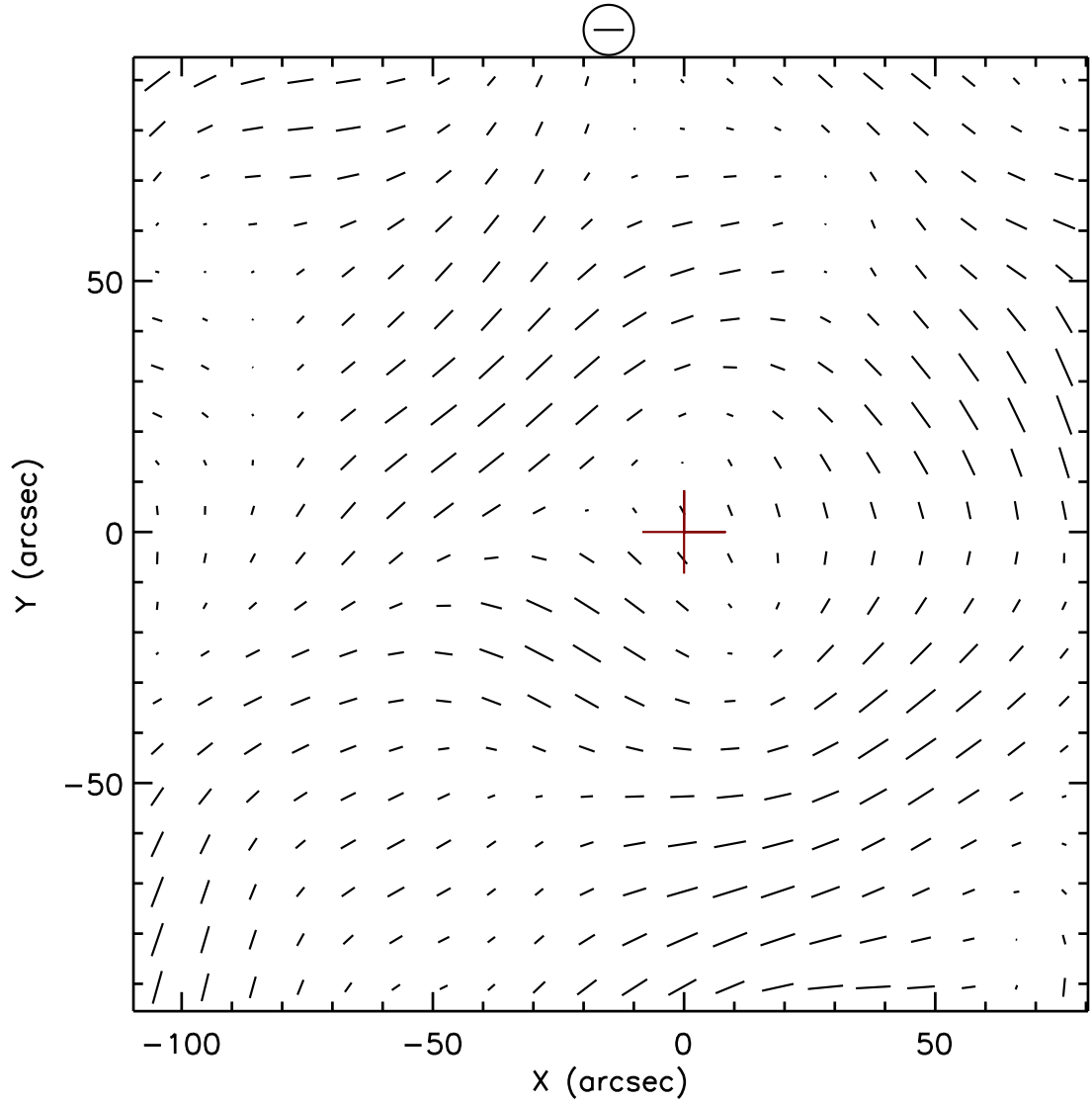}
\caption{Smoothed ellipticity distribution of background galaxies.
The ``whisker'' plot is created by smoothing the ellipticities
of the source population with a FWHM$\sim35\arcsec$ Gaussian kernel. The encircled stick just above the plot
shows the 10\% shear.
Tangential alignment of the sticks around
the cluster center (we mark the location of the BCG with the `+' symbol) is clear.}
\label{fig_whisker}
\end{figure}

Figure~\ref{fig_whisker} shows the so-called ``whisker'' plot, which displays the smoothed (with a FWHM$\sim35\arcsec$ Gaussian) ellipticity
distribution of the source galaxies. It is clear that the gravitational shear from the cluster
causes tangential alignments of the sticks near the cluster center. It is possible to attempt to
reconstruct a two-dimensional mass map directly from this smoothed ellipticity map with
the $g\sim\gamma$ approximation.
One of the straightforward implementations of the technique is the Fourier-space inversion method of 
Kaiser \& Squire (1993, hereafter KS93). As the KS93 algorithm is fast and easy to apply, it is still
extensively used by many authors although quite a few improvements have been suggested to minimize 
the artifacts in the KS93 method. One noteworthy problem is the spurious noisy structures
near the field boundary. This arises mainly because 1) signals are weak and biased near the boundary and 2)
masses outside the boundary can affect the shears inside. Seitz et al. (1998) showed that entropy-regularized mass 
reconstruction without smoothing galaxy shapes can 
suppress the boundary effect and also increase the resolution where the signal is strong.
In this paper, we use the mass reconstruction code of Jee et al. (2007a), which modified the method of Seitz et al. (1998)
so that strong-lensing data can be incorporated. Because we have not identified the multiple image systems nor
estimated photometric redshifts of individual galaxies, 
we proceed by turning off the strong-lensing capability of the code and also by assigning 
a single redshift of $z_{eff}=1.84$ to source galaxies (however, using the above relation $g\prime=(1+1.17 \kappa) g$ to account for the width
of the redshift distribution).
We refer readers to Jee et al. (2007a) for details of the algorithm. 

\begin{figure}
\plottwo{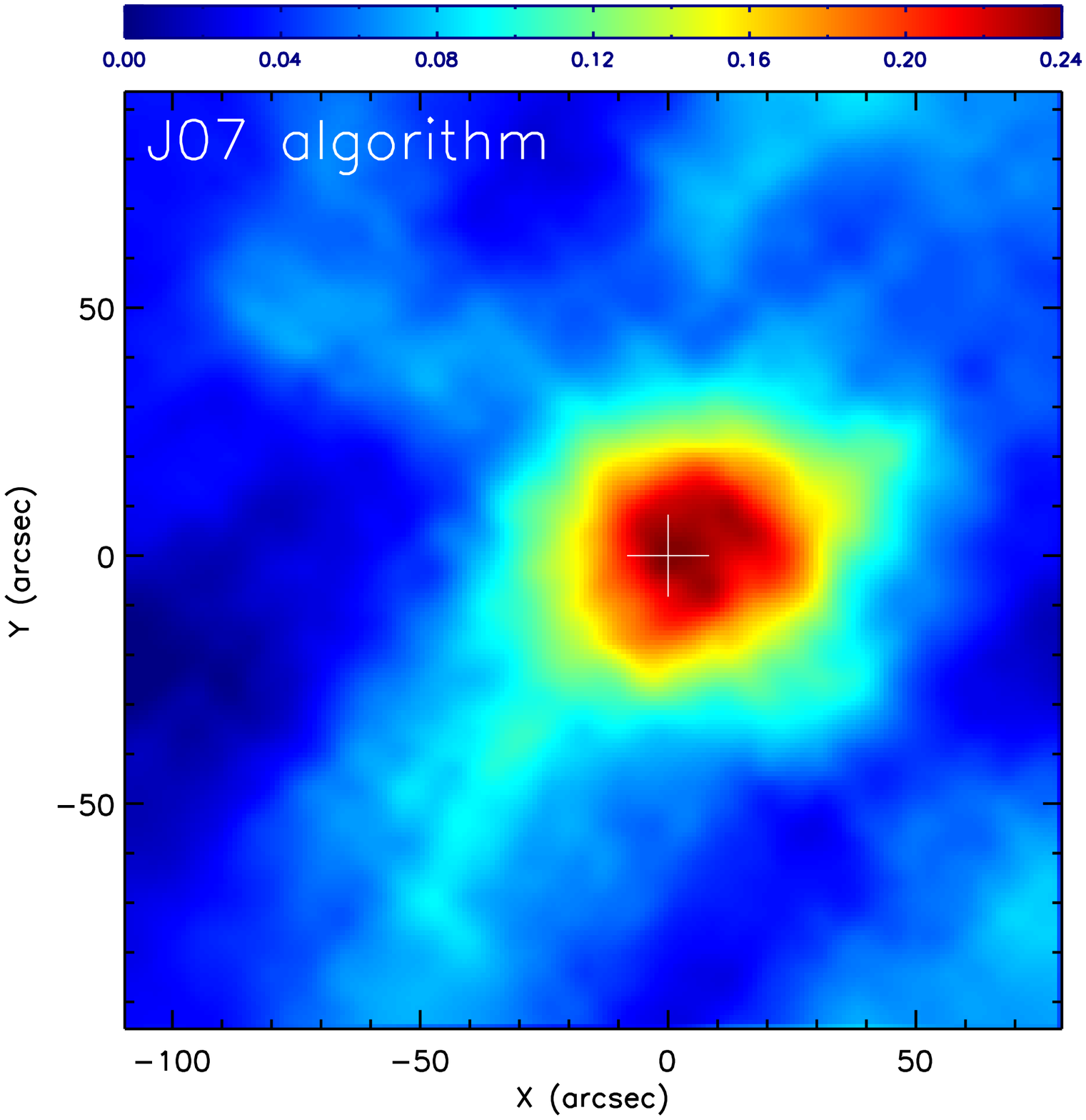}{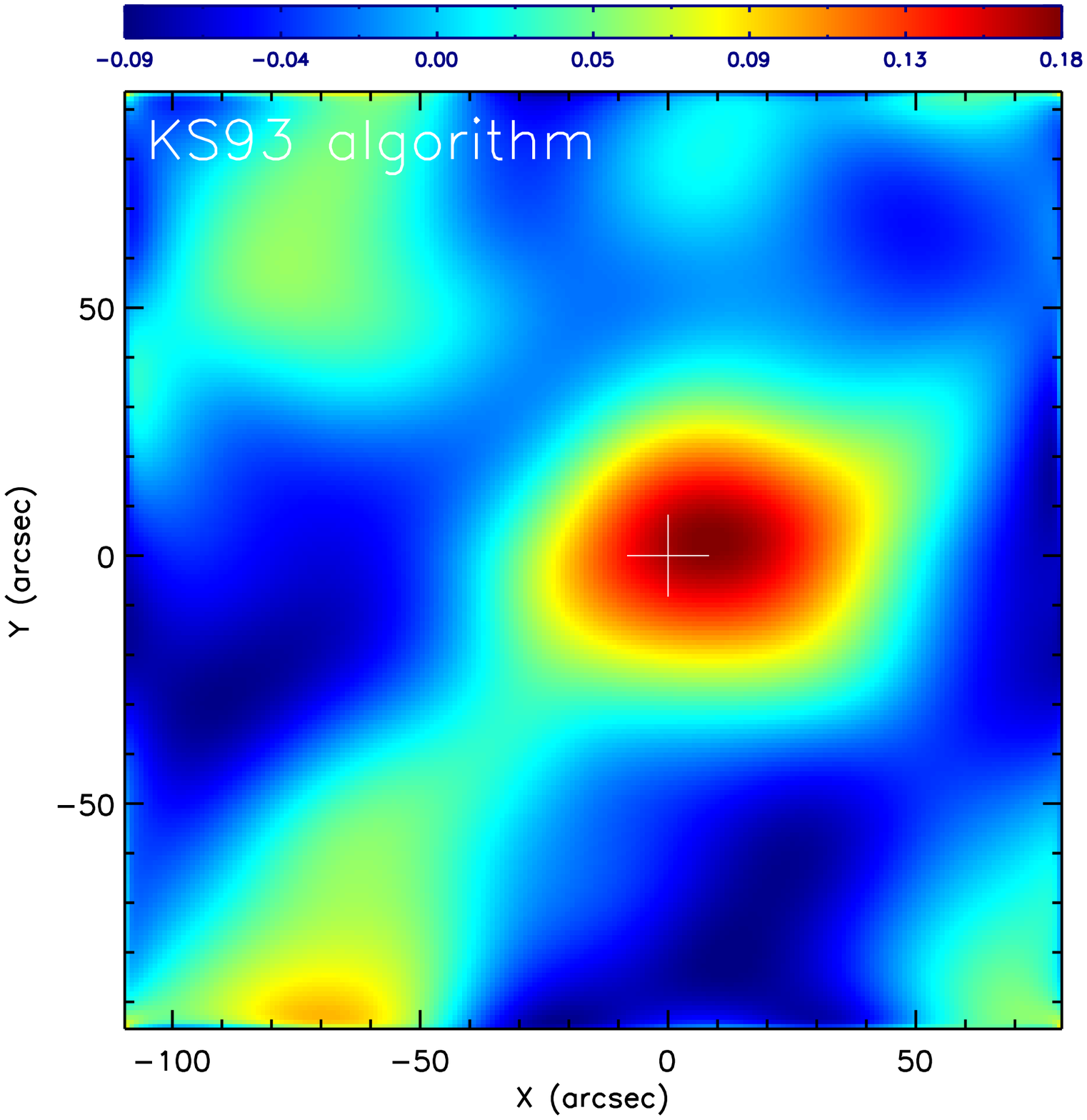}
\caption{Mass reconstruction of XMM2235. We use the maximum entropy regularization code of Jee et al. (2007) to
reconstruct the two-dimensional mass distribution of the cluster (left). The `+' symbol represents the location of the cluster BCGs.
The mass-sheet degeneracy is lifted using the NFW fitting result.
For comparison, we also display the KS93 mass reconstruction in the right panel. North is up and east is left.
\label{fig_j07_vs_ks93}}
\end{figure}

We display our mass reconstruction results in Figure~\ref{fig_j07_vs_ks93}.
The result obtained with the Jee et al. (2007a) code is shown in the left panel, and we show the
KS93 version in the right panel for comparison. Although both reconstruction methods clearly 
detect the cluster with high significance, the KS93 result displays substantial gratuitous substructures
near the field edges. These features have remarkably low significance in our maximum-entropy reconstruction, which
effectively employs a larger smoothing kernel for a weaker lensing signal (see Jee \& Tyson 2009 for a similar comparison). 
In addition, 
the method does not use the $g=\gamma$ approximation, which non-negligibly misinterprets the observed ellipticity
near the cluster center.

\begin{figure}
\plottwo{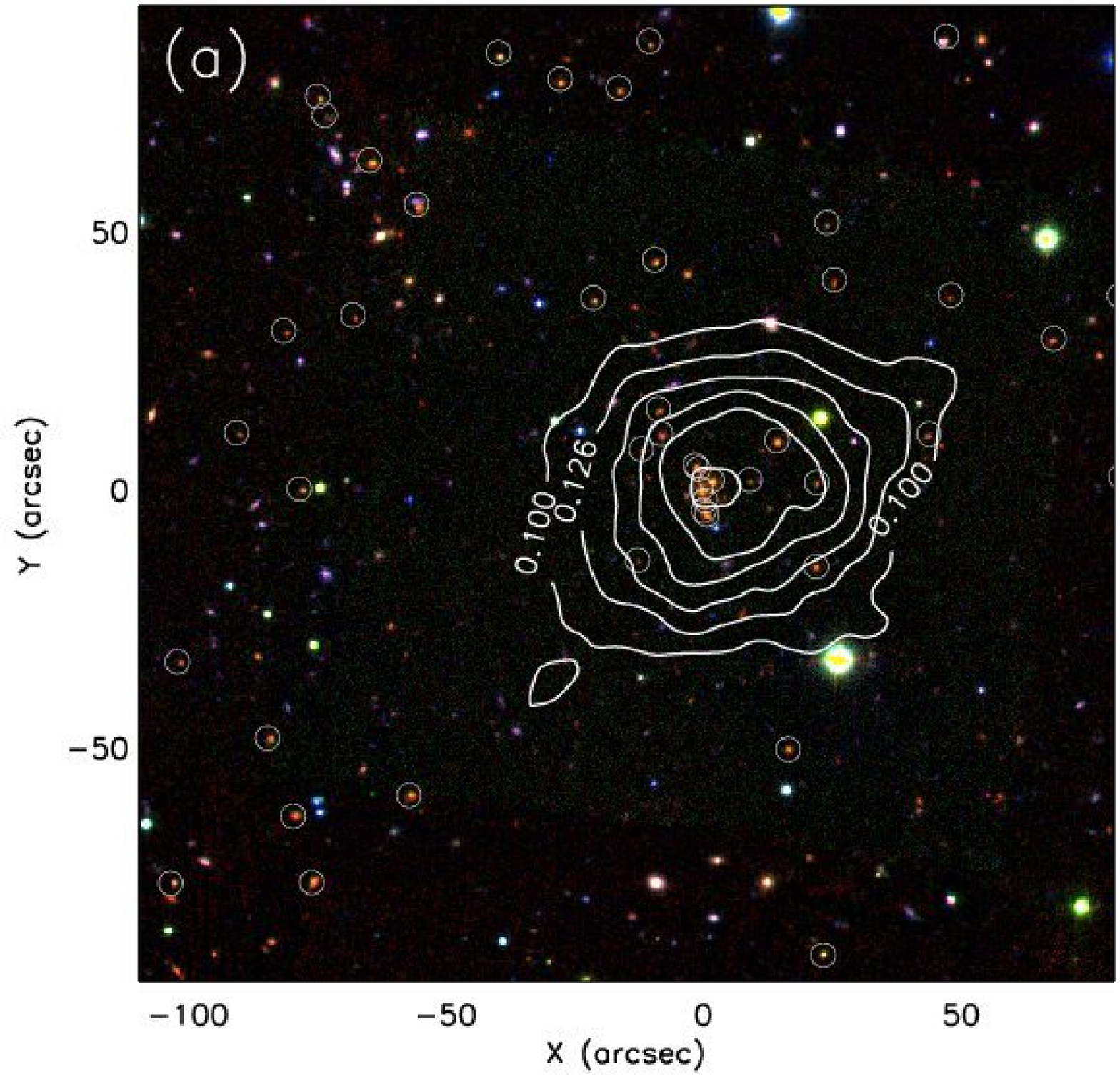}{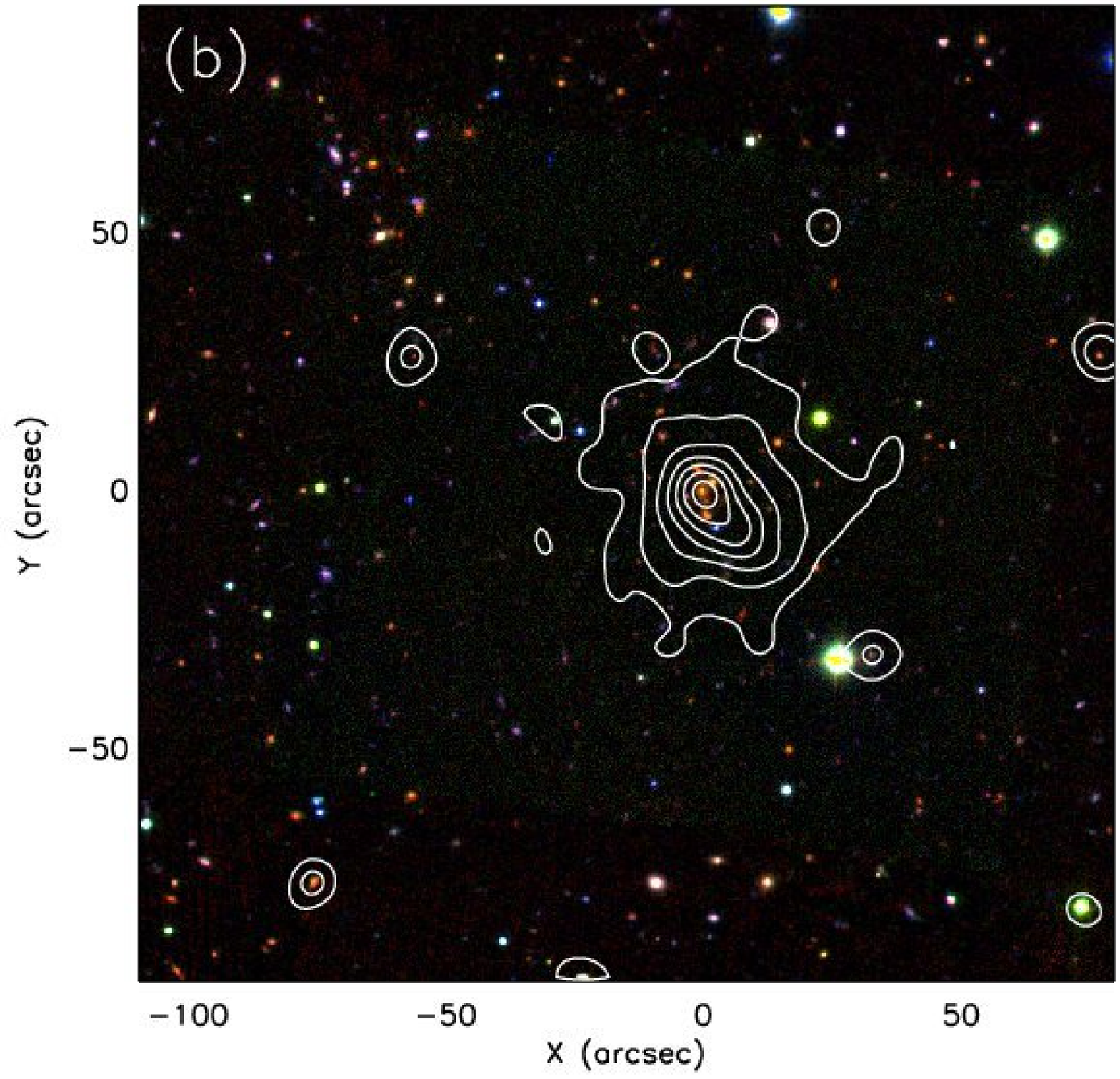}
\caption{Mass, X-ray, and galaxy comparison.
(a) The mass contours overlaid on top of the color composite image of the cluster field.
The red channel is an ISAAC $K_s$-band image, the green channel is a HAWK-I
J-band image, and
the blue channel is a FORS2
$R$-band image.
We denote the cluster red-sequence candidates (selected by $i_{775}-z_{850}$ colors) with circles.
(b) Chandra X-ray contours on top of the optical cluster image. The exposure corrected Chandra image is
smoothed with a FWHM$\sim7.4\arcsec$ Gaussian kernel. Contours are linearly spaced.
\label{fig_mass_vs_xray}}
\end{figure}

The mass peak is in excellent spatial agreement with the BCG and the X-ray peak (Figure~\ref{fig_mass_vs_xray}).
Together with the relaxed appearance of the mass and the gas, this agreement suggests that the cluster
is not undergoing any violent merger. However, we note that the global mass centroid lies
$\sim10\arcsec$ toward west from the BCG. Because the cluster is at a high redshift, the chance that the cluster
possesses any intervening foreground structure is high, and this potential foreground
structure without being very massive can affect the mass reconstruction. However, our study of the current spectroscopic
catalog of the field does not hint at this possibility. Moreover, the distribution of the red-sequence candidate galaxies
near the cluster center ($r\lesssim40\arcsec$) is in good agreement with the mass distribution.

\subsection{Tangential Shear and Mass Estimation\label{section_mass_estimation}}

\begin{figure}
\plotone{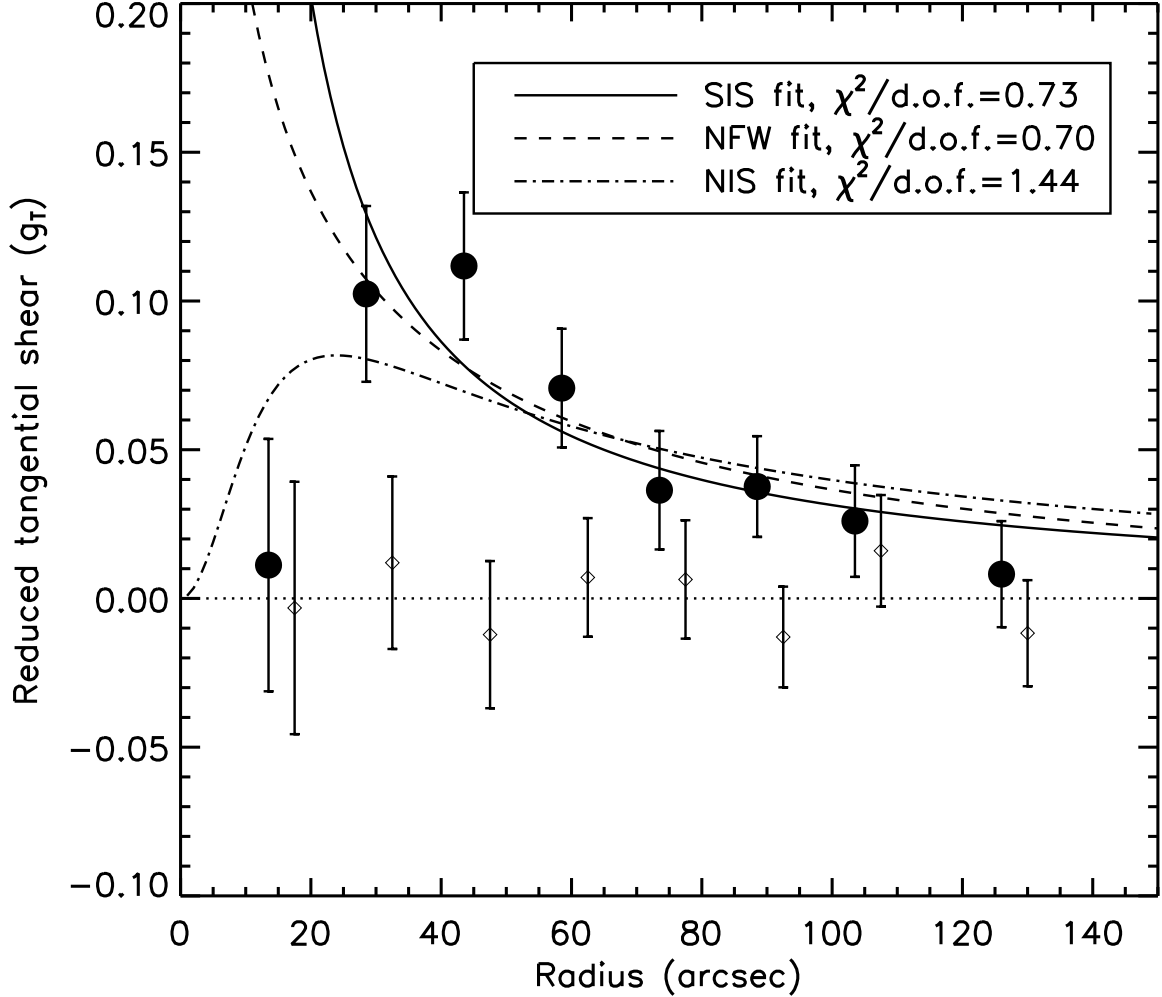}
\caption{Reduced tangential shear measured around XMM2235. The signal (filled circle) peaks around $r=30\arcsec-40\arcsec$ and
then decreases somewhat monotonically out to $r\sim140\arcsec$. The individual points are uncorrelated
and thus the statistical significance as a whole is very high ($>8\sigma$).
The diamond symbols represent the shears also measured from the same galaxies, however, with the galaxies
rotated by 45$\degr$. This component $<g_\times>$ must vanish as observed if the signal is indeed by lensing.
The dot-dashed line is the best-fit NIS model whereas the solid and dashed lines represent the 
best-fit SIS and NFW models, respectively. We did not use the innermost point ($r\sim15\arcsec$) for SIS and NFW fitting.
See the text for the summary of the fitting results.
\label{fig_tan_shear}}
\end{figure}

Adopting the location of the cluster BCG as the center of the cluster, we calculated the reduced tangential shears 
of XMM2235.3-2557 as a function of radial distance.
The filled circles in Figure~\ref{fig_tan_shear} clearly indicate that the cluster mass systematically
distorts the shapes of the source population out to $r\sim140\arcsec$ (at $r\gtrsim80\arcsec$
the annulus does not complete a circle). Because these data points are not correlated, the detection significance
is very high ($\gtrsim 8\sigma$).
The diamond symbols represent the results
of the so-called ``null'' test, which is measured in the same way except that the source galaxies
are this time rotated by 45$\degr$. Null tests are used to assess the amount of 
uncorrected systematics and the result in Figure~\ref{fig_tan_shear} supports that
no significant systematics are present. The error bars shown here only include the statistical
uncertainty set by the finite number of used galaxies. Hoekstra (2003) pointed out that 
background large scale structures are also important sources of uncertainty in cluster mass
estimation. We followed the formalism of Hoekstra (2003) and estimated this contribution.
Within the range $r<140\arcsec$, the induced uncertainty is  $\sigma_{\gamma}\sim0.01$. These values are
added in quadrature to the statistical uncertainty in our error analysis.

In fitting analytic mass profiles to the observed shear profile, 
we consider three parameterized models: singular isothermal sphere (SIS),
Navarro-Frenk-White (NFW) profile, and non-singular isothermal (NIS) sphere models.
The innermost data point at $r\sim10\arcsec$ being excluded, fitting a singular
isothermal sphere (SIS) model gives an Einstein radius of $\theta_E=6.05\arcsec\pm0.75\arcsec$ with 
$\chi^2/d.o.f=0.73$. This Einstein radius is valid for the effective redshift of the source population $z_{eff}=1.84$, and
thus it is interesting to examine if the value is consistent with
the location of the arc to the East side of the cluster BCG. The tentative spectroscopic
redshift of the object is $z\simeq3.3$ and the resulting Einstein radius increases to
$\theta_E (z=3.3)=14\arcsec\pm 2\arcsec$, which is in good agreement with the distance of $\sim12\arcsec$ between 
the arc and the BCG (the study of this strong lensing configuration with additional observations
is deferred to a forthcoming paper).
The SIS result predicts a velocity dispersion of the cluster as high as $1145\pm70~\mbox{km}~\mbox{s}^{-1}$, which
is somewhat larger (however, the error bars marginally overlap) than the value $762\pm265~\mbox{km}~\mbox{s}^{-1}$
measured from 12 spectroscopic members 
(Mullis et al. 2005). Nevertheless, this velocity dispersion that our SIS model predicts is
consistent with the X-ray temperature $8.6_{-1.2}^{+1.3}$ keV recently measured by $Chandra$ (Rosati et al. 2009). Under the
simplistic assumption of the energy equipartition (i.e., $\sigma_v=T^{1/2} /[0.59~m_p]^{1/2}$, where $m_p$ is the proton mass), the Chandra
temperature is translated into $1181\pm{86}~\mbox{km}~\mbox{s}^{-1}$.
Although the SIS assumption might lose its validity at large radii, these values are suggestive of
XMM2235 being a very massive system at $z=1.4$.

Blindly fitting an NFW profile does not well-constrain
two parameters of the model simultaneously. For example, concentration $c$ and scale
radius $r_s$ can trade with each other without largely altering the goodness of the fit whereas a pair
of parameters with a lower concentration give a higher mass.
This is a common problem in typical weak-lensing shear profile fitting if the
measurement does not extend to a sufficiently large distance. Nevertheless, 
the difficulty can be overcome by assuming a relation between the concentration parameter $c$
and the total cluster mass. One popular choice in the literature for the mass-concentration relation
is the result of Bullock et al. (2001), who showed that an average concentration decreases
with both halo mass and redshift according the following relation:
\begin{equation}
c_{vir}=\frac{9}{1+z} \left( \frac{M_{vir}}{8.12 \times 10^{14} ~h~ M_{\sun}} \right)^{-0.14}. \label{bullock2001}
\end{equation}
Using equation~\ref{bullock2001}, we found that the tangential shear profile
is best described by a halo with $c_{vir}=1.86\pm0.07$ and $M_{vir}=8.3_{-1.9}^{+2.6}\times10^{14}~M_{\sun}$
($\chi^2/d.o.f=1.1$)\footnote{$M_{vir}$ refers to the mass within a sphere, where
the mean density becomes $\sim166$ times the critical density $\rho_c$ at $z=1.4$ in the adopted cosmology whereas
$M_{200}$ is defined with the density contrast 200 $\rho_c$ regardless of cosmology and redshift. The
difference between $c_{vir}$ and $c_{200}$ is analogous.}. This virial mass is somewhat 
higher than what the above SIS fitting result and 
the X-ray temperature suggest, which motivates us to extend our experiments with different mass-concentration relations.

The strong dependence of the concentration parameter on halo mass and redshift implied by equation~\ref{bullock2001}
is shown to be in conflict with the result obtained from the Millenium Simulations (Springel et al. 2005) by
Gao et al. (2008). Their analysis of the relaxed clusters in the simulation supports much weaker dependence of
concentration parameter on halo mass and redshift for most massive clusters than is seen in Bullock et al. (2001).
If the relation claimed by Gao et al. (2008) is true, the concentration parameters predicted by equation~\ref{bullock2001}
are underestimated by $\sim35$\% at $z=1$ for $M\gtrsim3\times10^{14} M_{\sun}$, and the difference
increases for higher redshift. Noting the importance of the rigor of the uncertainty estimation
in the current study, we choose to include the scatter of the concentration parameter for simulated
clusters measured by Gao et al. (2008) in our estimation of the mass uncertainty.
By interpolating the results shown in Figure 5 and Table 1 of Gao et al. (2008), we adopt the range 
$c_{200}=3.20\pm0.75$ as the interval spanning the 68\% confidence limits for $M >5\times10^{14}~h~ M_{\sun}$
at $z=1.4$. By fitting an NFW model to the tangential shear by varying $c_{200}$
within the $c_{200}=3.20\pm0.75$ range, we found that the corresponding 1-$\sigma$ scatter in mass is $\sim11$\%,
which is similar to the $\sim11$\% uncertainty caused by the source redshift estimation error and
the $\sim14$\% uncertainty caused by both statistical noise and cosmic shear.
Adding these errors in quadrature, we estimate that the spherical mass within $r_{200}=1.1$ Mpc is 
$M_{200}=(7.3\pm1.3)\times10^{14} M_{\sun}$ ($\chi^2/d.o.f=0.70$).

Using an NIS [$\kappa=\kappa_0/(r^2+r_c^2)^{1/2}$] model yields a result indistinguishable from the SIS result
as long as the innermost point is excluded. Including the point gives
a core radius of $r_c=14\arcsec\pm5\arcsec$ and a normalization of $\kappa_0=4.86\pm1.19$ ($\chi^2/d.o.f=1.4$).

\begin{figure}
\plotone{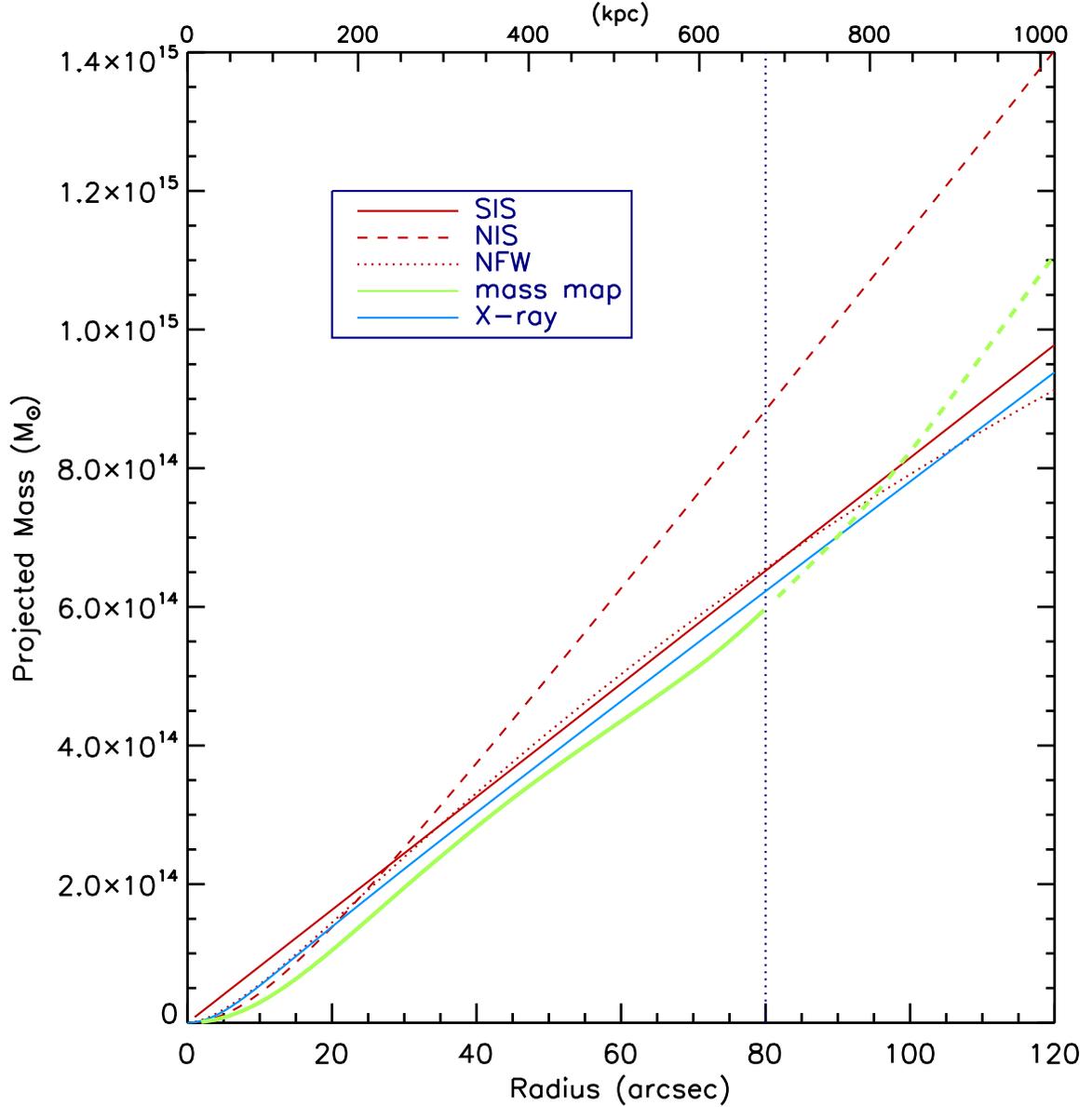}
\caption{Projected mass profiles of XMM2235 from the various methods used in this paper. The NIS fitting
to the tangential shears gives significantly higher masses than the other methods at large radii. However,
we do not consider the result as representative of the entire cluster mass profile.
The direct use of the mass map gives consistent values with the SIS and NFW results at $r\lesssim80\arcsec$ 
(see the vertical dotted line);
at $r\gtrsim80\arcsec$ the data comes from a limited azimuthal range (green dashed line), and the bias
in mass estimation increases with radius.
The uncertainties, which we omit in the plot for readability, are about 12\%, 16\%, and 14\% 
for the SIS, NFW, and mass map results, respectively at $r=1$ Mpc. Note that these percentage errors
change slightly with radius for the NFW and mass map results. We did not
include the independent error of $\sim11$\% due to the uncertainty in $z_{eff}$. 
The X-ray mass is based on an isothermal $\beta$ model with $T=8.6$ keV, $r_c=10.7\arcsec$, and $\beta_X=0.61$ (Rosati et al. 2009).
This is in good agreement with our lensing results.
\label{fig_mass_comparison}}
\end{figure}

We compare in Figure~\ref{fig_mass_comparison} the projected mass profiles that these three models
predict. At $r\lesssim80\arcsec$, the difference between the SIS and the NFW results is insignificant and
the mass grows almost linearly with the radius in both cases. However, due to the asymptotic
$\rho(r)\propto r^{-3}$ behavior, the growth of the NFW mass profile becomes slower at large radii.
The NIS assumption leads to substantially higher masses than what the other two models predict ($\sim45$\%
higher than the NFW result at $r=80\arcsec$!), and we do not consider this model as representative of
the XMM2235 mass profile.

As already discussed in \textsection\ref{section_theory}, the convergence map that we obtained in 
\textsection\ref{section_mass_reconstruction} can be used to estimate the mass when we lift
the mass-sheet degeneracy utilizing this parametric cluster mass description. For this experiment we constrained 
the $60\arcsec<r<80\arcsec$ annulus to have the mean $\kappa$ set by the NFW result; the
SIS result gives a similar value in this annulus. The thick
green line in Figure~\ref{fig_mass_comparison} shows the mass profile obtained from this
convegence map.
Because we cannot obtain a full azimuthal range of data at $r>80\arcsec$ (marked by the vertical dotted line), the bias
in the mass estimation increases with radius (thick dashed green line).
With this caveat, we estimate that the total projected mass
within 1 Mpc is $(1.03\pm0.16)\times10^{15}~M_{\sun}$.

Rosati et al. (2009) report that the recent analysis of the $Chandra$ data yields
an X-ray temperature of $8.6_{-1.2}^{+1.3}$ keV, a core radius
of $r_c=10.7\arcsec$, and a $\beta_X$ index of 0.61. 
The blue line in Figure~\ref{fig_mass_comparison} illustrates the 2-D projected 
cluster mass profile based on a single isothermal $\beta$ model estimated with 
these parameters. This X-ray mass profile is highly consistent with the lensing
results.

Although a single isothermal $\beta$ model has been considered
an inadequate representation in a number of recent studies, we stress that for this
particular case the displayed X-ray mass profile
is in a good agreement with our full model-independent, X-ray surface brightness 
deprojection result. This model-independent profile closely resembles the 
isothermal beta model profile 
out to $r\sim~0.5$ Mpc, gradually deviates from the isothermal 
$\beta$ model result at $r > 0.5$ Mpc, and
gives a $\sim11$\% lower value at $r=1$ Mpc (however, the statistical 
uncertainty of the X-ray result is already at the $\sim20$\% level, and
the X-ray emission is significant only out to $\sim500$ kpc).
This model-independent, X-ray surface brightness deprojection result will
be presented in a forthcoming paper.

\section{How Rare is an XMM2235-like cluster at $z=1.4$?}

Our weak-lensing analysis shows that XMM2235 is indeed a massive cluster at $z=1.4$.
The projected mass of the cluster is $\sim8.5\times10^{14}~M_{\sun}$ within 1 Mpc.
Such a massive cluster is extremely rare when the age of
the universe is about a third of its current value, and thus it is interesting to quantify
the probability of finding an XMM2235-like cluster within the survey volume containing XMM2235.

One of the most well-tested halo mass functions is the so-called universal mass function of
Jenkins et al. (2001). From extensive studies of $N$-body simulation results, they found that
the mass function takes a universal form regardless of cosmology, epoch, and the power spectrum
as long as the distribution is expressed in terms of $\mbox{ln}~ \sigma^{-1}$, where $\sigma(M,z)$
is the linear density field rms for a given mass scale $M$ at a redshift of $z$.
The universal mass function is approximated by
\begin{equation}
f(\sigma)=\frac{M}{\rho_0} \frac{\mbox{d}~n(M,z)} { \mbox{d ln}~\sigma^{-1} } = A~\mbox{exp}(-|\ln \sigma^{-1} + B|^{\epsilon}),
\end{equation}
\noindent
where $n(M,z)$ is the number density of halos with mass less than $M$ at $z$ and $\rho_0$ is 
the mean density of the universe.
Jenkins et al. (2001) demonstrated that for the $\Lambda$CDM model, when a halo mass is defined as 
the total mass within the sphere that encloses 324 times
the mean density of the universe, the parameters $A=0.316$, $B=0.67$, and $\epsilon=3.82$ can describe
the numerical results in the range $-0.7< \mbox{ln}\sigma^{-1} < 1.0$ with $20$ \% accuracy. 

\begin{figure}
\plotone{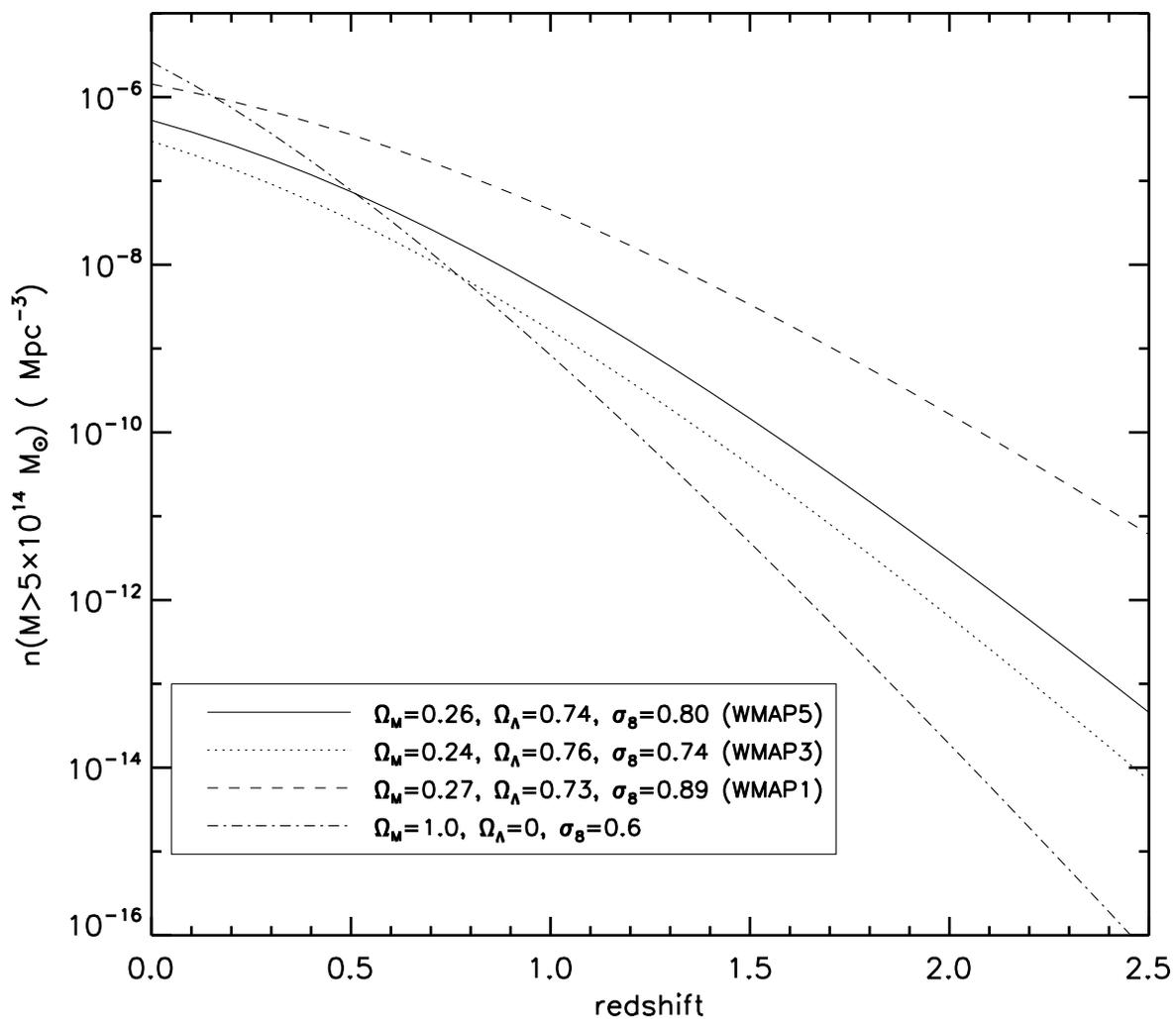}
\caption{Evolution of comoving number density of massive clusters for different cosmological parameters.
We use the Jenkins et al. (2001) fitting formula for the spherical overdensity group finder (324 times the mean
density of the universe) to produce the plot. 
\label{fig_mass_function_evolution}}
\end{figure}

Using the above halo abundance function,
we present in Figure~\ref{fig_mass_function_evolution} the redshift evolution of comoving number density of clusters whose
masses are greater than $5\times10^{14} M_{\sun}$ for various cosmological parameters.
This $5\times10^{14} M_{\sun}$ threshold mass is calculated using the NFW model, which gives the lowest mass for XMM2235 
among the presented results.
At the redshift of the cluster the spherical overdensity of 324 is reached at $r=0.93$ Mpc ($\sim110\arcsec$). 
The spherical volume encloses a total mass of $(6.4\pm1.2)\times10^{14}~ M_{\sun}$, where the error bar includes 
the statistical uncertainty, the effect of possible interloping background structures, the scatter in concentration
parameter, and
the uncertainty of the mean redshift of the background galaxies.
Hence, the used threshold $5\times10^{14} M_{\sun}$ 
is the conservatively chosen  1-$\sigma$
lower limit of our measurement.

The old $\tau$CDM cosmology (dot-dashed) predicts the fastest evolution of the massive halo abundance 
(nearly 6 orders-of-magnitude difference between $z=0$ and 1.4). 
This is why previously the existence of the massive cluster MS1504-0321 at $z=0.84$ has been argued as evidence for $\Omega_M <<1$ (e.g., Bahcall \& Fan 1998).
For cases with a non-zero cosmological constant, we consider the cosmological parameters derived from
Wilkinson Microwave Anisotropy Probe 1-, 3-, and 5-year results (referred to as WMAP1, WMAP3, and WMAP5, respectively).
The WMAP3 parameters (Spergel et al. 2007) underestimate the expected cluster abundance significantly
(by nearly one order of magnitude at $z=0$ and two orders of magnitude at $z=1.4$) with
respect to the WMAP1 (Spergel et al. 2003) results. The difference at $z=0$ was recently viewed as problematic
because the earlier (WMAP1) results were consistent with the observed local cluster abundance 
(e.g., Yeps et al. 2007; Evrard et al. 2008); however see Rosati et al. (2002)
and references therein for the cluster mass function evolution studies that favor low-normalization.
Interestingly, the recent WMAP5 parameters (Dunkley et al. 2008) give values somewhere between the two sets of parameters and 
somewhat relieve the tension. Rines et al. (2008) however noted that the WMAP5
result best matches the observation when still a velocity segregation of 1.13 (the ratio of the velocity
dispersion of the galaxy to dark matter) is assumed, which implies that without the velocity segregation more shift toward the WMAP1 result
is favored by the observed cluster abundance; nevertheless, they claimed that the combination of WMAP5 with supernovae and baryonic
acoustic oscillation results removes this residual.

Although our analysis is confined to a single cluster, the surprisingly high mass of XMM2235 provides an
important opportunity to extend the aforementioned test to the $z\gtrsim1.4$ regime.
XMM2235 was discovered in an archival XMM-Newton observation during the pilot study phase (Mullis et al. 2005; 
Boehringer et al. 2006). 
A field was
selected if it had a clean exposure time above 10 ksec without large-scale X-ray sources.
The median flux limit of the fields is $\sim10^{-14}$ $\mbox{erg}~\mbox{s}^{-1}~\mbox{cm}^{-2}$ in the 0.5-2 keV band, and this
gives a maximum redshift of $z_{max}\sim 2.2$,
where a massive ($M>5\times10^{14}~M_{\sun}$) cluster can be detected.
The total survey area $11~\mbox{deg}^2$ then corresponds to a comoving volume 
of $V(1.4<z<2.2)=10^8~\mbox{Mpc}^3$; the optical identification process of the cluster
cluster catalog at $z<1.4$ in the XDCP survey is
not complete, and thus here we choose the redshift lower limit to be $z_{min}=1.4$ assuming
that this is the only cluster we detected at $z\gtrsim1.4$ with mass in excess of $M=5\times10^{14}~M_{\sun}$.
How many XMM2235-like clusters are expected to be found within this volume?
The theoretical value can be estimated by evaluating the volume-weighted
average of the abundance in Figure~\ref{fig_mass_function_evolution}
within the $1.4<z<2.2$ redshift interval.
Adopting the WMAP5 cosmological parameters, 
we obtain $n=5.1\times10^{-11}~\mbox{Mpc}^{-3}$. Therefore, we expect
to observe only $N\sim5\times10^{-3}$ clusters in the survey, which
leads us to conclude that the discovery of XMM2235 is 
a rare event with a probability of $\lesssim 1$\%.

\section{SUMMARY AND CONCLUSIONS}
We have presented a weak-lensing analysis of the galaxy cluster XMM2235 at $z\simeq1.4$
using the HST/ACS data. Despite the high redshift of the lens, the signal is
clear both in the one-dimensional tangential shear profile and in the two-dimensional mass reconstruction.
This clear detection is enabled in part by the high mass of the cluster, which
is estimated to be $\sim8.5\times10^{14}~M_{\sun}$ within a $r=1$ Mpc aperture.
Because the spectroscopic survey of the field does not hint at any significant foreground
cluster along the line-of-sight, we attribute this lensing mass to the cluster XMM2235. The X-ray temperature
$8.6_{-1.2}^{+1.3}$ keV of the cluster recently measured by $Chandra$ predicts a consistent mass with our lensing
result.

This high mass of the cluster is unusual at $z=1.4$ in the current hierarchical structure formation paradigm for commonly
accepted cosmological parameters. Using the Jenkins et al. (2001) mass function while
choosing the 1-$\sigma$ lower limit as a threshold mass, we
estimate that the expected number of XMM2235-like clusters within the survey volume
is only $N\sim5\times10^{-3}$.
Therefore, the discovery of the cluster is certainly
a rare event with a probability of $\lesssim 1$\%.

Mindful of the restrictions in the interpretation of this $single$ event,
there are some recent studies, which hint at the possibility
that the discovery of XMM2235 might not be a statistical outlier. Fedeli et al. (2008) 
demonstrate that the observed strong-lensing statistics is incompatible with the 
WMAP cosmological parameters, favoring $\sigma_8 \gtrsim 0.9$. 
If the survey volume containing XMM2235 represents the mean property of the universe, and
we simplistically
attribute this discrepancy to the potential underestimation of the matter density fluctuation,
the discovery of XMM2235 would certainly point to a higher value of $\sigma_8$.
For example, a $\sigma_8$ value of 0.9 (i.e., $2.5 ~\sigma$ upper limit
of the WMAP5 result) would predict that there are $\sim5\times10^{-2}$ such clusters within 
the survey, alleviating the tension by a factor of 10. 

Obviously, increasing $\sigma_8$ is not the only way to explain the existence of the cluster
within the survey. Considering in particular that a number of previous studies using X-ray clusters
for cosmology have favored a low $\sigma_8$ (0.7-0.8) value (Rosati et al. 2002 and references
therein), we should be also open to other scenarios, wherein more massive clusters form at higher
redshifts.
Broadhurst \& Barkana (2008)
found that the Einstein radii of the well-known clusters are a factor of two larger than the
predicted values, suggesting that the observed concentration is higher than expected.
Within the paradigm that a halo's concentration reflects the mean density of
the universe at the time of its formation, the claimed discrepancy suggests that
clusters may form earlier than we currently believe.
Fedeli \& Bartelmann (2007) claim that the presence of the early dark energy can increase
$\sigma_8$ on non-linear scales, which can allow more clusters to form earlier and thus
more efficiently produce massive clusters at high redshift. 
Another scenario that accomodates the existence of XMM2235 within the survey is the non-Gaussian primordial
density fluctuation (Peebles 1999).
The non-Gaussian density fluctuation has been shown to increase the abundance of galaxy clusters
at the high end (e.g., Willick 2000; Mathis et al. 2004; Sadeh et al. 2007; Grossi et al. 2007), producing
more rare, massive clusters at higher redshift.

Finally, we consider the possibility that the employed mass function in the current
study is highly biased. Jenkins et al. (2001) states that their universal function
matches the simulation results within 20\%. Using gas-dynamic simulations, Stanek
et al. (2009) claim that inclusion of baryons systematically causes $\sim30$\%
deviation with respect to the results from the dark matter only simulation. 
Unfortunately, because the direction of the shift depends on the simulation methods, 
it is not yet clear toward which direction the true mass function should move.
Let us think of an extreme case, where the real mass function should be 50\% higher.
If we use the $2-\sigma$ lower limit (i.e., $M_{thr}=4\times10^{14}~M_{\sun}$ as 
opposed to $M_{thr}=5\times10^{14}~M_{\sun}$ used above) of
the mass estimate as a threshould mass, we expect to detect $3\times10^{-2}$ such clusters
within the survey. 

Currently, we are undertaking weak-lensing analysis of more than 20 $z\gtrsim 1$
clusters based on HST/ACS images (M. Jee et al. in preparation). The study
will increase the number of $z>1$ clusters whose masses are measured through weak-lensing
by a factor of 10. Although the result does not provide a mass function at $z>1$,
a similar study using the most massive clusters in the sample will provide us with an
opportunity to judge whether or not XMM2235 is a outlier.
Therefore, extensive discussions on the implication of the surprisingly high mass
of the cluster are deferred to our future publications. 

M. J. Jee acknowledges support for the current research from the TABASGO foundation presented in the form of
the Large Synoptic Survey Telescope Cosmology Fellowship.
Financial support for this work was in part provided by NASA through program GO-10496 and GO-10698
from the Space Telescope Science Institute, which is operated by AURA,
Inc., under NASA contract NAS 5-26555. This work was also supported
in part by the Director, Office of Science, Office of High Energy and
Nuclear Physics, of the U.S. Department of Energy under Contract
No. AC02-05CH11231. We thank Stefano Borgani and Hu Zhan for verifying our estimation
of the cluster abundance.

\clearpage

\appendix
\section{CHARGE TRANSFER INEFFICIENCY STUDY FROM STELLAR PHOTOMETRY \label{section_appendix_cti}}

We cannot find any explicit report of the turnaround of the CTI for low-count objects that we discuss in 
\textsection\ref{section_cti_correction}
elsewhere in the literature, and therefore it is worthwhile here to investigate if the effect is also observed in stellar
photometry, which has been a conventional approach in the CTI measurement. The absence of the CTI turnaround
in stellar photometry does not necessarily falsify the claim in \textsection\ref{section_cti_correction}
because many factors can conspire in such a way that the effect is detected only in ellipticity measurement, but
is unobservable in photometry. For example, if the trails following bright sources are concentrated in the first few pixels 
while the trails following faint sources are spread over a much larger area (i.e., flux-dependent
time constant), the CTI turnaround
may appear in ellipticity measurement, but absent in aperture photometry. Nevertheless, independent detection
of the feature from photometry will certainly support the existence of the CTI turnaround.

We retrieved the HST CTI monitoring datasets 10043, 10368, and 10730, which observe the globular cluster NGC104 between 
the years 2004-2006. For parallel CTI studies, two different points apart by $\sim1.5\arcmin$ ($\sim2000$ pixels) 
are used so that an object undergoes two different charge transfers in readout (see Riess \& Mack 2004 for
the details of the observing strategy). 
We detected stars using the DAOFIND algorithm on individually drizzled images, and 
performed 3-pixel radius aperture photometry. We measured the local sky value from an $r=20-30$ pixel annulus.
We agree with Riess \& Mack (2004) that choosing to use local sky values introduces higher statistical noise
than the global sky measurement scheme, but prevents any systematics arising from sky gradients or
residual flat-fielding errors. 
We inverse-transformed the location of a star into the CCD coordinate, and this allows us to
compute the difference in the amount of the charge transfer between exposures. 

Not surprisingly, the CTI study from this photometry method gives much less stringent constraints
on the CTI slope than the CR-based study (\textsection\ref{section_cti_correction}), not only
because the number of data points is small, but also because time-dependent
PSF variation can affect the photometry within the 3-pixel radius aperture. Nevertheless, we detect
the mitigation of the CTI effect for low-count objects also in this photometry experiment as follows.

\begin{figure}
\plottwo{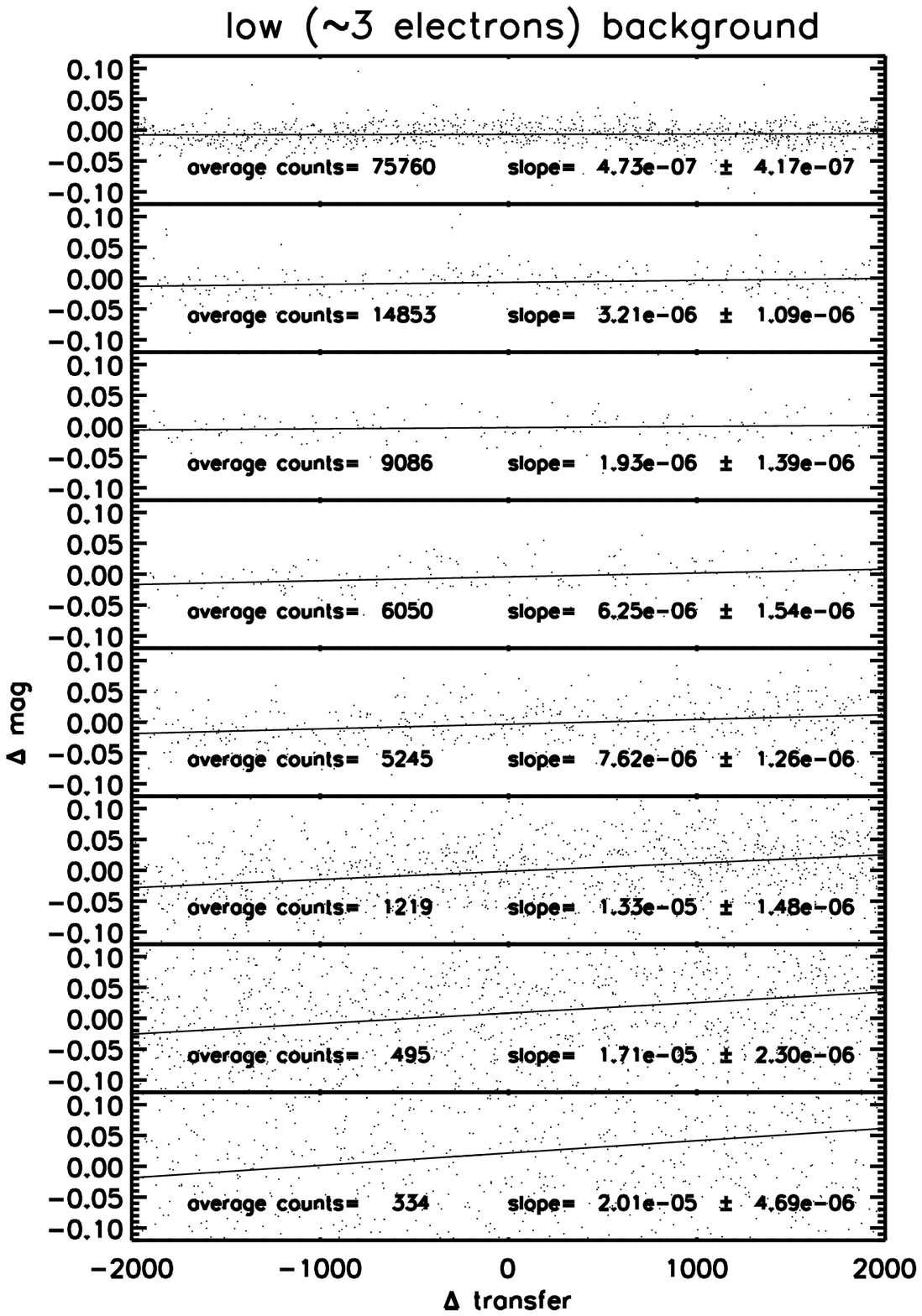}{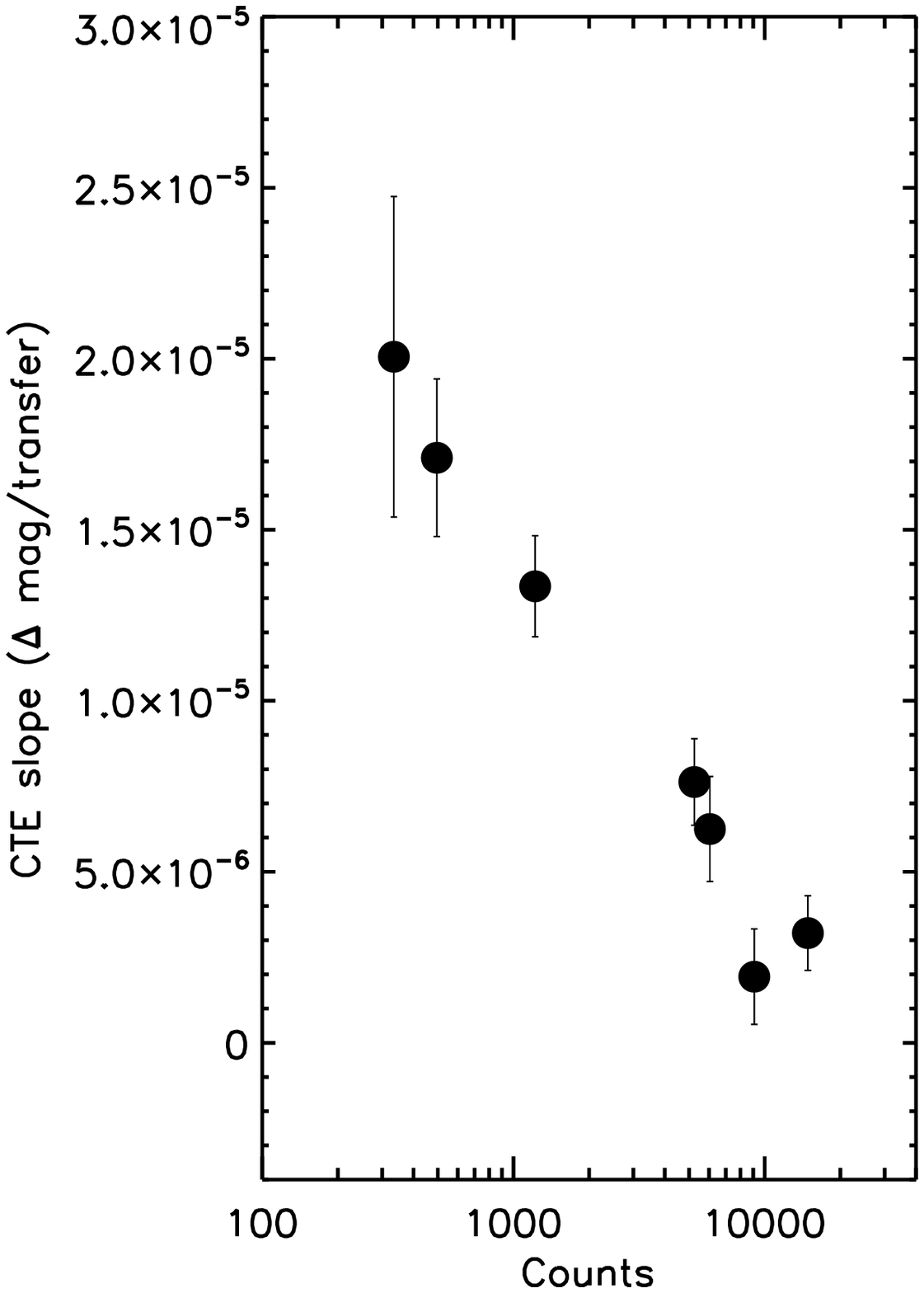}
\caption{CTI study from low background images via aperture photometry.
The left-hand-side plots show $\Delta$ mag versus $\Delta$ transfer with
each panel showing the result for different fluxes. We observe in this example
where the background count is low that the CTI degrades as fluxes decrease.
The righ-hand-side plot displays the CTI slope as a function of mean flux (counts).
\label{fig_cte_low_bg_photometry}}
\end{figure}

\begin{figure}
\plotone{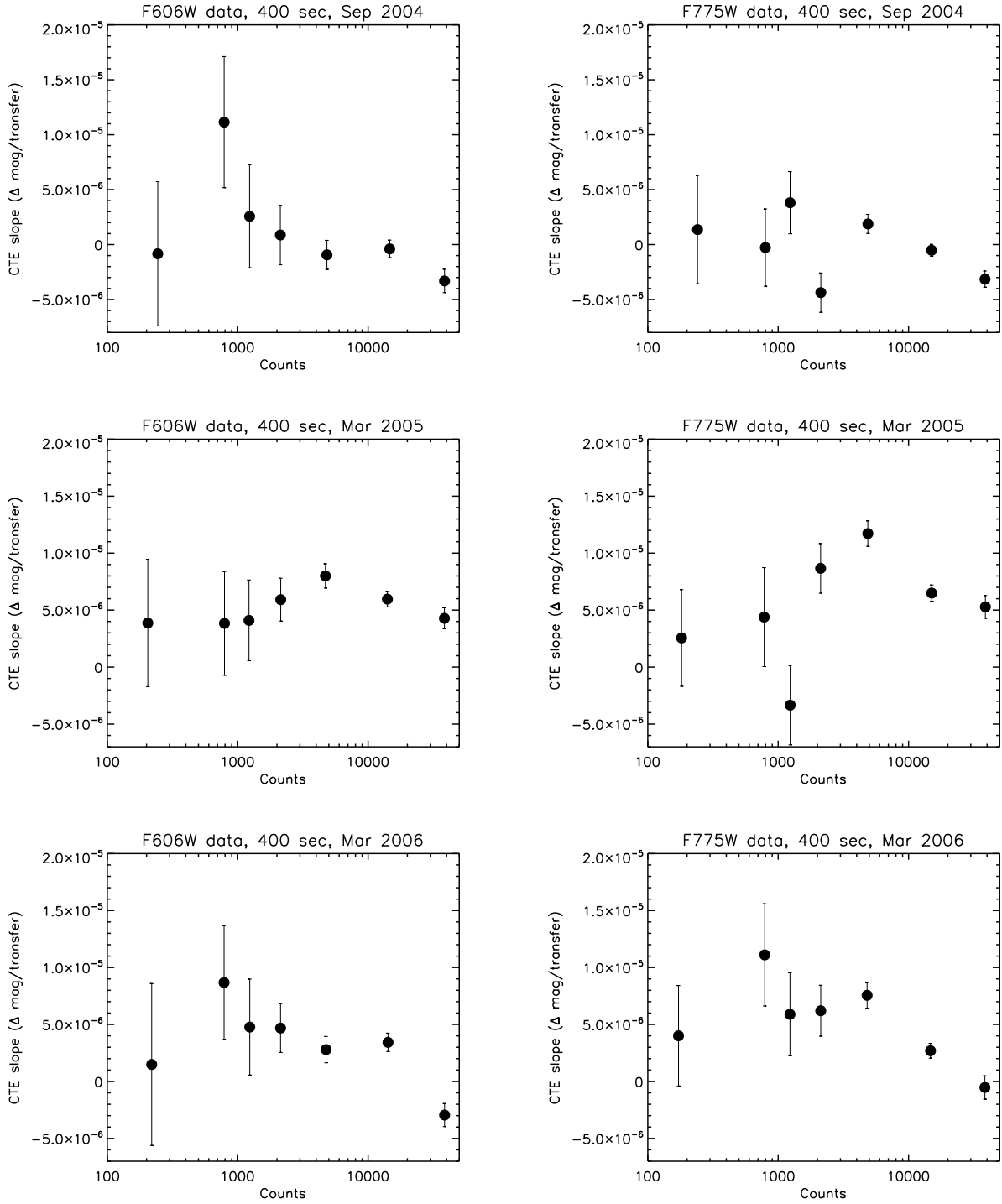}
\caption{CTI study from high background images via aperture photometry.
The background counts in these images are relatively high, 
ranging from 30 $\mbox{e}^{-}$ to 40 $\mbox{e}^{-}$.
Although the significance in each image is low, 
the flux losses due to CTI are always suppressed at low counts ($\sim 200~\mbox{e}^{-}$). 
We interpret this observation as supporting the existence of the CTI turnaround 
that we independently detect with much higher S/N in the cosmic ray test 
(\textsection\ref{section_cti_correction}).
The CTI degradation does not appear to be a simple function of time
in this test.
We suspect that our 3-pixel aperture photometry may be also affected by time-dependent PSF variation.
\label{fig_cte_high_bg_photometry}}
\end{figure}

Figure~\ref{fig_cte_low_bg_photometry} shows the flux loss of objects due to the CTI effect measured from
two low background images taken in September 2004.
In plotting $\Delta$ magnitude versus $\Delta$ transfer ($left$), we
follow the convention of Riess \& Mack (2004), where a negative $\Delta$ transfer means
that the amount of the readout transfer is less in the first image. Therefore, if charges
are lost during transfer, we expect to see positive $\Delta$ m for positive $\Delta$ transfer
as observed. The different panels represent the results for different fluxes. We observe
the gradual steepening of the CTI slope as the mean flux decreases (see also the right panel),
which is in accordance with the conventional understanding of CTI properties. The exposure
time is only 30 s, and the mean background count is $\sim3~\mbox{e}^{-}$. For this low level
of background, we do not expect to observe the CTI mitigation for low-count objects as seen here.
However, this montonic increase of CTI is not observed when we repeat the experiment
with relatively high-background images. In Figure~\ref{fig_cte_high_bg_photometry} we show the 
relation between CTI slope and flux for
six pairs of observations taken in the years 2004, 2005, and 2006 for the F606W and F775W filters.
The mean background of these images ranges from 30 $\mbox{e}^{-}$ to 40 $\mbox{e}^{-}$. Based
on the CR-based study results in \textsection\ref{section_cti_correction}, we expect that
for this level of background the turnaround happens at $200-300$ $\mbox{e}^{-}$.
Although the significance in each pair of images is low, we notice that
the flux losses due to CTI are always suppressed at the expected location ($200-300$ $\mbox{e}^{-}$).
Because we do not observe this pattern in other pairs of short exposure images,
we interpret this CTI suppression at low flux ends as supporting our claim of the CTI turnaround that is
alternatively and more significantly seen in our cosmic ray test.

\end{document}